\documentclass[12pt]{article}
\usepackage{cite,epsfig,amssymb,amsmath,amstext,  graphicx,color,subfigure,xcolor,mathrsfs,tikz}
\evensidemargin \oddsidemargin

\numberwithin{equation}{section}

\usetikzlibrary{decorations.pathmorphing}
\usepackage{latexsym}
\usepackage{hyperref}
\usepackage{bbm}
\usepackage{color}

\hypersetup{colorlinks,%
  citecolor=blue,%
  linkcolor=red,%
  pdftex}
\usepackage[numbers,sort&compress,square]{natbib}
\usepackage{varioref}
\usepackage{slashed}
\usepackage[normalem]{ulem}
\usepackage[makeroom]{cancel}
\usepackage{pdfsync}
\usepackage{mathrsfs}
\usepackage{tikz}
\usepackage{pgfplots}

\begin{document}

\vspace{5mm}

\begin{center}

{{\Large \bf 

Quantum Inhomogeneous Field Theory: Unruh-Like Effects and Bubble Wall Friction}}
\\[5mm]

Jeongwon Ho$^1$, ~~O-Kab Kwon$^1$, ~~Sang-Heon Yi$^2$  \\[2mm]
{\it $^1$Department of Physics,~Institute of Basic Science, Sungkyunkwan University, Suwon 16419, Korea} \\
{\it $^2$ Physics Department $\&$ Natural Science Research Institute University of Seoul, Seoul 02504, Korea} 

{\it ( freejwho@gmail.com, ~okab@skku.edu, ~~shyi704@uos.ac.kr )}
\end{center}
\vspace{15mm}

\thispagestyle{empty}

\begin{abstract}
\noindent

In this paper, we study a free scalar field in a specific (1+1)-dimensional curved spacetime.  By introducing an algebraic state that is locally Hadamard, we derive the renormalized Wightman function and explicitly calculate the covariantly conserved quantum energy-momentum tensor up to a relevant order.
From this result, we show that the Hadamard renormalization scheme, which has been effective in traditional quantum field theory in curved spacetime, is also applicable in the quantum inhomogeneous field theory.  As applications of this framework, we show the existence of an Unruh-like effect for an observer slightly out of  the right asymptotic region, as well as the vanishing of quantum frictional effect in the leading order ($e^{- bx}$) on the bubble wall expansion during the electroweak phase transition in the early universe.

\end{abstract}


\vskip 1.0in

\hrulefill

All authors contributed equally.

 

 \newpage

\section{Introduction}
Quantum field theory (QFT) has been very successful for describing the nature, yielding very precise match between theoretical predictions and experimental tests. Based on this success, it is natural to explore possible extension of this framework to more general cases. However,  early attempts to extend QFT on  curved spacetime have encountered  various conceptual and computational difficulties with some ambiguities in their formulation. Nevertheless,   the limit of the Newton constant, $G_{N}\rightarrow 0$  is believed to be well-described by a semi-classical approximation. In this approximation, the curved background is taken as a given configuration  while various fields, including graviton field, in the background are considered as  fluctuating quantum fields. In this case, the absence of Poincar\'e symmetry requires a more careful treatment of the relevant Hilbert space and field operators.  Mathematically rigorous and conceptually superior approach in this direction is known as algebraic approach~\cite{Haag:1992hx, Wald:1995yp, Hollands:2014eia,Khavkine:2014mta,Witten:2021jzq}.

Another interesting system without Poincar\'e symmetry would be field theory with spacetime dependent mass and couplings in flat spacetime. 
To distinguish the standard field theory with Poincar\'e invariance, we designate field theory with position-dependent mass and couplings, as {\it inhomogeneous field theory} (IFT).  There have been various motivations and origins to arrive at this kind of field theories. One of the most frequently encountered situation is the existence of some classical background configuration and regarding this configuration as a given fixed one.  IFT provides us an interesting model for various physical situations. For instance,   the bubble wall model in the cosmological setup can be regarded as a concrete example of 
IFT~\cite{Polyakov:1974ek, Coleman:1977py,Callan:1977pt,Linde:1981zj,Moore:1995ua,Bodeker:2017cim,Azatov:2020ufh,Kubota:2024wgx}.  These IFTs  have also been supersymmetrized with either non-abelian~\cite{Kim:2018qle, Kim:2019kns, Arav:2020obl,Kim:2020jrs} or abelian~\cite{Hook:2013yda, Tong:2013iqa, Adam:2019yst,Kim:2023abp, Kim:2024gpu, Kim:2024gfn,Jeon:2024jbs} gauge groups in (1+2) or (1+3) dimensions. See also for supersymmetric IFT models in (1+1) dimensions~\cite{Kwon:2021flc, Ho:2022omx}. Although classical aspects of IFT may be  interesting  in their own rights, the quantum effects or quantum aspects of IFT would be more relevant in some regimes. Therefore, it would be quite interesting direction to explore the quantization of IFT. However, the absence of Poincar\'e invariance in IFT leads to various difficulties,  similar  to quantum field theory on curved spacetime (QFTCS).

Since there is no complete consensus on the quantization of IFT, we have proposed a quantization scheme for specific $(1+1)$-dimensional IFT in~\cite{Ho:2022omx,Kwon:2022fhv}. Concretely, this approach is based on a simple observation that $(1+1)$-dimensional classical field theory on curved spacetime (FTCS) in a specific gauge can be rewritten as  $(1+1)$-dimensional IFT in the classical context. By using the classical conversion, one may adopt the same methodology, {\it i.e.  algebraic approach} of QFTCS to IFT. Interestingly, this proposal can be implemented in a very concrete way, at least, for `free' field theories, which is worked out in~\cite{Ho:2022omx,Kwon:2022fhv}.  As a next step of this approach, we explore the conversion from the vacuum expectation value (VEV) of energy-momentum tensor in QFTCS to {\it quantum inhomogeneous field theory} (QIFT) with a specifically chosen state in this paper.

The energy-momentum tensor is one of crucial quantities for understanding physical systems. In particular,  it has been a primary focus in semiclassical quantum field theory on curved spacetime. There has been a standard procedure for obtaining the energy-momentum tensor and its VEV in Poincaré-invariant theories. It is typically deduced as a density of the spacetime translation generators (with some improvement procedure known as Belifante tensor). However, it is complicated when the background exists even in  Poincaré-invariant theories. For instance, in the Landau level problem, one can introduce two types of momenta —canonical and kinematical—  with their corresponding energy-momentum tensors. The choice of momentum depends on the specific interpretation or application, even at the classical level. At the quantum level, selecting an appropriate operator ordering with renormalization is also of importance in this problem. Since the product of field operators is not well-defined at coincident spacetime points, we must also take a more careful approach to define the energy-momentum tensor operator and compute its VEV.  One complication stems from the distributional  nature of field operators~\cite{Hollands:2009bke,Hollands:2023txn,Witten:2023qsv}. All these difficulties are bypassed by using the normal ordering and the choice of the Poincar\'e invariant global vacuum in Minkowski field theories with Poincar\'e symmetry.

On the other hand, defining a useful and meaningful energy-momentum tensor in other situations may be challenging, partially due to the lack of global or local symmetries.  Though in generally covariant theories, the classical energy-momentum tensor is obtained as a response of the action to metric variation, 
it becomes complicated in the context of quantum field theory. In addition, it  becomes more subtle to compute the VEV of  energy-momentum tensor operator  in curved spacetime coming from the intricate ``vacuum'' structure of FTCS.  One approach to address this  is to relax the conditions required for the vacuum state in Minkowski spacetime and to focus on the algebraic structure of the field operators, introducing the relevant states at later stage. 
It is noteworthy that IFT also provides a context where all these issues arise,  necessitating new insights for their resolution.

 In scalar IFT  with  a varying  mass in space, where the Poincar\'e symmetry is broken, no preferred vacuum  exists, contrary to quantum field theory on  Minkowski spacetime with the preferred  Poincar\'e invariant vacuum.  In such case, to define a `vacuum' and calculate the renormalized two-point function for that vacuum state is complicated. Though algebraic approach may be taken without introducing a preferred  `vacuum'  in IFT,  a reasonable subtraction scheme should be taken explicitly to compute the VEV of renormalized energy-momentum tensor.  However, currently in QIFT, such a scheme is not established or there is no consensus on the correct approach.  On the other hand, in QFTCS, the well-known Hadamard method leads to  a reasonable renormalized two-point function, and  furthermore the VEV of the energy-momentum tensor $\langle T_{\mu\nu}\rangle_{{\rm H}}$, defined for the Hadamard state, which is  covariantly conserved~\cite{Wald:1978pj, Birrell:1982ix,Kay:1988mu, Moretti:2001qh}.  
 
In~\cite{Ho:2022omx}, we have proposed  how to relate QFTCS and QIFT, focusing on (1+1) dimensions, based on the simple fact that a FTCS action in a given background metric can be converted classically to an IFT action. We have argued that this can give a new perspective on the quantization of IFT. In this regard, it is expected that  the Hadamard renormalization in QFTCS may provide a guiding principle for defining the renormalization of IFT two-point function in (1+1) dimensions.  In this paper, by using the Hadamard renormalization method for two-point function of FTCS, we identify the renormalized two-point function in QIFT and construct a `covariantly conserved' VEV of energy-momentum tensor in IFT  two-point function, and through this, we aim to interpret quantum effects  in QIFT.

Especially, in this paper, we  focus on `free' scalar field theory with  the position-dependently varyingly mass in $(1+1)$-dimensional flat space.
The scalar field we consider starts as massless at $ x \rightarrow -\infty$ and monotonically increases to approach a constant mass at  $x \rightarrow +\infty$. 
This simple scalar IFT  may  have various physical applications. For instance, in the early universe,  we can model the Higgs condensate bubble expansion as a `free' scalar IFT  where the mass of the scalar field varies continuously  along the bubble wall.  The quantum effects in  this model could give us interesting physical implication of the bubble expansion, as was explored for the two-point function in~\cite{Kubota:2024wgx}. Along this line, we compute the VEV of the energy-momentum tensor in a specific IFT, building on our previous proposal~\cite{Ho:2022omx}.

Concretely, we consider the following  action of a free massive scalar field in a specific curved spacetime $ ({\cal M}, g_{ab}) $, 
\begin{align}
S_{\text{FTCS}} = \int_{\mathcal{M}} d^2 x \sqrt{-g} \Big( -\frac{1}{2} g^{\mu\nu} \nabla_\mu \phi \nabla_\nu \phi - \frac{1}{2} m_0^2 \phi^2 - \frac{1}{2} \xi \mathcal{R} \phi^2 \Big),
\end{align}
where $m_0$ is a mass parameter, $\xi$ a dimensionless parameter, and $\mathcal{R}$ the curvature scalar of the background metric. 
The background curved spacetime $ ({\cal M}, g_{ab})$ is not an arbitrary one but a specifically chosen by the supersymmetry requirement in our setup and therefore, the corresponding mass function is quite constrained.  Then, the above scalar field action on ${\cal M}$  is converted to the scalar IFT action as 
\begin{equation} \label{IFTact}
S_{\text{IFT}} = \int d^2x \Big[ -\frac{1}{2} \eta^{\mu\nu} \partial_{\mu} \phi \partial_{\nu} \phi - \frac{1}{2} m_{{\rm eff}}^2(x) \phi^2 \Big]\,,
\end{equation}
which is our main interest in this paper. This action has been used for  the bubble wall model in~\cite{Kubota:2024wgx}. Specifically, we compute the pressure difference between slightly out of the Higgs phase region and the Higgs phase. At the end of the day, we show that the quantum effect of the vacuum for the observer who located near the Higgs phase lead to the Unrhu-like effect and the vanishing of the frictional force in the leading order ($e^{- b x}$) against the bubble wall expansion.

This paper is organized as follows. In section~\ref{sec2}, we review the supersymmetric formulation for our background and its corresponding IFT with canonical quantization. In this section, we also provide mode solutions and their corresponding various vacua. In section~\ref{sec3}, after carefully introducing the `vacuum' state for the observer residing slightly out of the asymptotic right region, we present the renormalized two-point functions for our model by adopting the so-called Hadamard renormalization method in the context of QFTCS. By employing the improved point-splitting method  associated  with the covariant   energy-momentum tensor~\cite{Moretti:2001qh},  we present our results on the VEV of energy-momentum tensor in our background in section~\ref{sec4}.  We provide concrete values of the VEV of energy-momentum tensor, consistent with the flat Minkowski limit.  In section~\ref{sec5}, we provide IFT interpretation of our results in section~\ref{sec5}. Based on these results, we present the Unrhu-like effect and the vanishing of the frictional quantum effects of our vacuum in the leading order $(e^{- b x})$ on the expansion of the bubble wall that occurs during the electroweak phase transition in the early universe. In the final section, we give some comments on our results and future directions.

\section{Supersymmetric Curved Background}\label{sec2}

In this section, we  briefly summarize the construction of the  (1+1) dimensional supersymmetric  model of FTCS discussed in the previous paper~\cite{Ho:2022omx}. Introducing the supersymmetric background metric and the analytic mode solution for the scalar field $\phi$,  we  review the canonical quantization of the scalar field $\phi$ in the context of the supersymmetric background as well.

\subsection{SUSY background in (1+1) dimensions}

In \cite{Ho:2022omx}, a supersymmetric field theory on a curved spacetime background (SFTCS) was constructed, whose action is given by
\begin{align}\label{SFTCST}
S_{{\rm SFTCS}} = \int d^2 x \sqrt{-g} &\Bigg[ -\frac{1}{2} g^{\mu\nu} \nabla_\mu\phi \nabla_\nu\phi + \frac{i}{2} \bar{\Psi} \gamma^\mu \nabla_\mu \Psi + \frac{i}{2} \left(\frac{\partial^2 \mathcal{W}}{\partial \phi^2}\right) \bar{\Psi} \Psi - \frac{1}{2} \left(\frac{\partial \mathcal{W}}{\partial \phi}\right)^2 
\nonumber\\
&~ - f(\mathcal{R}) \mathcal{W}(\phi, \mathcal{R}) - \mathcal{U}(\phi, \mathcal{R}) \Bigg],
\end{align}
where $\phi$ and $\Psi$ represent a real scalar field and a two-component Majorana spinor, respectively, and $\mathcal{R}$ denotes the curvature scalar of the background metric.
Here, $\mathcal{W}$ and $\mathcal{U}$ are functions of $\phi$ and $\mathcal{R}$, given by
\begin{align}
\mathcal{W}(\phi, \mathcal{R}) &= \sum_{n \ge 1} \mathcal{F}_n (\mathcal{R}) \phi^n, 
\nonumber\\
\mathcal{U}(\phi, \mathcal{R}) &= \sum_{n \ge 1} \mathcal{G}_n (\mathcal{R}) \phi^n,
\end{align}
where $f$, $\mathcal{F}_n$, and $\mathcal{G}_n$ are functions of $\mathcal{R}$ that ensure the covariance of the action \eqref{SFTCST}. Under the supersymmetric variation,
\begin{align}
\delta \phi &= i \bar{\Psi} \epsilon, \nonumber\\
\delta \Psi &= - \gamma^\mu \nabla_\mu \phi \epsilon + \left( \frac{\partial \mathcal{W}}{\partial \phi} \right) \epsilon,
\end{align} 
the action \eqref{SFTCST} remains invariant, provided that the supersymmetric parameter $\epsilon$ satisfies the generalized Killing spinor equation
\begin{align}\label{gkse}
\nabla_\mu \epsilon = \frac{1}{2} f \gamma_\mu \epsilon,
\end{align}
along with the relation $\sqrt{-g} \, \mathcal{G}_n^2 = - 4 \partial_{+} \mathcal{F}_n \partial_{-} \mathcal{F}_n$, where $\partial_{\pm} \equiv \frac{1}{2} \left(\frac{\partial}{\partial t} \pm \frac{\partial}{\partial x}\right)$. 
In the cases of flat and AdS$_2$ spacetimes, the function $f$ becomes a constant, resulting in two supersymmetries corresponding to $\mathcal{N} = (1,1)$. However, when $f$ is not a constant and the Ricci scalar $\mathcal{R}$ depends only on the spatial coordinate $x$, there exists a single supersymmetry \cite{Kwon:2021flc}. For further details, refer to \cite{Ho:2022omx}.

To solve the generalized Killing spinor equation \eqref{gkse}, we introduce the (1+1) dimensional metric in the conformal gauge as
\begin{align}\label{conmet}
ds^2 = e^{2 \Omega(t,x)} \left( - dt^2 + dx^2 \right).
\end{align}
It was shown in \cite{Ho:2022omx} that there is no supersymmetric solution for a time-dependent $\Omega$ that satisfies the generalized Killing spinor equation \eqref{gkse}. Therefore, we consider the case where $\Omega = \Omega(x)$, which allows for a single supersymmetry. In this case, the equation \eqref{gkse} reduces to the relation
\begin{align}\label{f(R)}
f(\mathcal{R}) = e^{- \Omega} \Omega',
\end{align}
where $' \equiv \frac{\partial}{\partial x}$.
Specifically, for the case where $f$ is a linear function of $\mathcal{R}$, such as 
\[
f(\mathcal{R}) = \frac{\xi}{m_0} \mathcal{R},
\]
where $\xi$ is a dimensionless parameter and $m_0$ is a constant parameter with dimension one, the relation \eqref{f(R)} can be rewritten as a differential equation for $\Omega(x)$:
\begin{align}
\Omega'' + \frac{m_0}{2 \xi}  \Omega' e^\Omega= 0.
\end{align} 
A solution to this differential equation is given by 
\begin{align}\label{susybac}
e^{\Omega(x)} = \frac{1}{a + e^{-b x}},
\end{align}
where $a$ and $b$ are integration constants that satisfy $ab = \frac{m_0}{2 \xi}\ne 0$. Then, the metric is given by 
\begin{align}\label{ourmet}
ds^2 = \frac{1}{(a + e^{-bx})^2}\left( -dt^2 + dx^2\right).
\end{align} 
Without loss of generality, one can choose $b > 0$ by using the reflection symmetry of the $x$-coordinate.
The curvature scalar of the metric given by \eqref{ourmet} is
\begin{align}\label{R}
\mathcal{R} = 2 a b^2 e^{-bx},
\end{align} 
which indicates that the geometry exhibits a naked null curvature singularity at $x \to -\infty$. As shown in \cite{Ho:2022omx}, this singularity is  \textit{mild} in the sense that wave propagation remains well-posed in this background.
Various supersymmetric field models can be put on this background. For instance, the sine-Gordon model, among others, can be formulated in this setting.

\subsection{Free SFTCS and supersymmetric IFT (SIFT)}\label{fSFSI}

As a specific example of  SFTCS with \( f(\mathcal{R}) = \frac{\xi}{m_0} \mathcal{R} \), we consider a free theory with the following choice of superpotentials:
\begin{align}
\mathcal{W}(\phi, \mathcal{R}) &= \frac{1}{2} m_0 \phi^2, 
\nonumber\\
\mathcal{U}(\phi, \mathcal{R}) &= 0.
\end{align}
Substituting these into \eqref{SFTCST}, the action \( S_{{\rm SFTCS}} \) becomes 
\begin{align}\label{SFTCSTf}
S_{{\rm SFTCS}} = \int d^2 x \sqrt{-g} \Big[ -\frac{1}{2} g^{\mu\nu} \nabla_\mu \phi \nabla_\nu \phi + \frac{i}{2} \bar{\Psi} \gamma^\mu \nabla_\mu \Psi + \frac{i}{2} m_0 \bar{\Psi} \Psi 
- \frac{1}{2} m_0^2 \phi^2 
- \xi \mathcal{R} \phi^2 \Big].
\end{align}
This model is supersymmetric under the supersymmetric variations
\begin{align}
&\delta \phi = i \bar{\Psi} \epsilon,
\nonumber\\
&\delta \Psi = - \gamma^\mu \nabla_\mu \phi \epsilon + m_0 \phi \epsilon,
\end{align}
on the supersymmetric background \eqref{susybac}. The equation of motion for \(\phi\) is given by
\begin{align}\label{FTCSTeom}
(-\Box + m_0^2 + \xi \mathcal{R} ) \phi = 0, \qquad \Box = \frac{1}{\sqrt{-g}} \partial_{\mu}\left( \sqrt{-g} g^{\mu\nu}\partial_{\nu} \right).
\end{align}
Conversely, one can say that we have supersymmetrized the above bosonic equation in (1+1) dimensions, resulting in the supersymmetric action \eqref{SFTCSTf}.

Note that we can rewrite the equation of motion \eqref{FTCSTeom} as 
\begin{align}\label{SIFTeom}
\left(-\partial^2 + m_{{\rm eff}}^2(x) \right) \phi = 0, \qquad \partial^2 = \eta^{\mu\nu} \partial_{\mu} \partial_{\nu},
\end{align} 
where \( m_{{\rm eff}}^2(x) = e^{2 \Omega}( m_0^2 + \xi \mathcal{R}) \). This equation represents the equation of motion for the inhomogeneous scalar field \(\phi\) with a position-dependent mass \( m_{{\rm eff}}(x) \). 
The corresponding SIFT action has the form \cite{Ho:2022omx}
\begin{align}\label{SIFTf}
S_{{\rm SIFT}} = \int d^2 x \Big[ -\frac{1}{2} \eta^{\mu\nu} \partial_{\mu} \phi \partial_{\nu} \phi + \frac{i}{2} \bar{\psi} \gamma^\mu_{{\rm F}} \partial_{\mu} \psi + \frac{i}{2} m(x) \bar{\psi} \psi - \frac{1}{2} m_{{\rm eff}}^2(x) \phi^2 \Big],
\end{align}
where \(\psi = e^{\frac{\Omega(x)}{2}} \Psi\), \(\gamma_{{\rm F}}^\mu\) are the  gamma matrices in the flat spacetime, and \(m(x)\) is related to \(m_{{\rm eff}}(x)\) by 
\begin{align}\label{meffs}
m_{{\rm eff}}^2(x) = m^2(x) + m'(x), \quad   
m(x) = m_0 e^{\Omega(x)}.
\end{align}
It should be noted that since  general covariance, which is a fundamental requirement in gravitational theory, is absent in the context of SIFT, all the physics in the SFTCS action \eqref{SFTCSTf} for different gauges can not be captured by the single SIFT action  \eqref{SIFTf}. What we have done is to convert the SFTCS  action in a selected conformal gauge to the corresponding SIFT action. 
Therefore, the algebraic structures of quantum fields, including their causal structures, of FTCS and IFT (even in the absence of supersymmetry) are identical. Based on this observation, we proposed in~\cite{Kwon:2022fhv} the application of the algebraic methodology developed in QFTCS to QIFT. This approach aligns with perspectives from researchers  on algebraic quantum field theory, particularly in the context of CCR algebra (see, for example, Remark 8-(4) on page 14 of \cite{Khavkine:2014mta}).

Now, we solve the Klein-Gordon equation with the space-dependent mass given by \eqref{meffs}. Since the SIFT Lagrangian in \eqref{SIFTf} possesses time translation symmetry, the mode solution for the scalar field, $u(\pmb{x})$, with a frequency $\omega$ can be expressed as
\begin{align}\label{uw}
u(\pmb{x}) \equiv \frac{1}{\sqrt{2\omega}} e^{-i\omega t} \phi_\omega(x).
\end{align}
Substituting this into the equation of motion \eqref{SIFTeom}, we obtain the equation for $\phi_\omega$:
\begin{align}\label{Aphiw}
A \phi_\omega(x) = \omega^2 \phi_\omega(x), \qquad  A = -\frac{d^2}{dx^2} + m_{\rm eff}^2(x).
\end{align}
The differential operator $A$ can be rewritten as
\begin{align}\label{ADD}
A = D_- D_+, \qquad D_{\pm} = \pm \frac{d}{dx} - m(x),
\end{align}
which confirms the positivity of the symmetric operator $A$.

Consequently, the equation \eqref{Aphiw} can be interpreted as a Schr\"{o}dinger equation within the framework of supersymmetric quantum mechanics (SQM) \cite{Cooper:1994eh}, where the factored operator $A$ in \eqref{ADD} corresponds to the Hamiltonian in SQM, with the superpotential identified as $W_{\rm QM}(x) = -m(x)$. Using SQM techniques and the factored form of the operator $A$ in \eqref{ADD}, we can define two partner Hamiltonians, $A^{(1)} \equiv A = D_-D_+$ and $A^{(2)} = D_+D_-$, along with the corresponding potentials $V^{(1)} = m_{\rm eff}^2 = W_{\rm QM}^2  - \frac{dW_{\rm QM}}{dx}$ and $V^{(2)} = W_{\rm QM}^2 + \frac{dW_{\rm QM}}{dx}$, respectively.

For our metric \eqref{ourmet}, we have \begin{align}\label{meff2}
m_{{\rm eff}}^2(x) = \frac{(m_0^2 e^{bx} + 2 \xi a b^2 )e^{bx}}{(a e^{bx} + 1)^2}.
\end{align}
As seen from the form of the potential $V^{(1)}(x) = m_{\rm eff}^2(x)$ in \eqref{meff2}, there are two sets of potentials determined by the sign of the solution parameter $a$, which is related to other parameters by $ab = \frac{m_0}{2 \xi}$.  For $a  < 0$, the geometry is divided by $x_* = - \frac{1}{b} \ln |a|$ into two asymmetric regions, $-\infty <x < x_*$ and $x_* <x < \infty $. In this scenario, the potential in SQM falls under the category of Eckart potentials. However, since we are interested in discussing the geometry defined over the entire spatial domain, $-\infty < x < \infty$, we focus on the case $a > 0$. 
The two partner potentials \( V^{(1)} \) and \( V^{(2)} \) in SQM are given by  
\begin{align} \label{RMpot}
V^{(1)}(x) &= 2b^{2}\xi^{2} \left[1  +  \tanh\left(\frac{b}{2}(x-x_{0})\right) \right] - \frac{b^{2}\xi}{2} (2\xi - 1) \frac{1}{\cosh^{2}\left(\frac{b}{2}(x-x_{0})\right)}, 
\nonumber\\
V^{(2)}(x) &= 2b^{2}\xi^{2} \left[1  +  \tanh\left(\frac{b}{2}(x-x_{0})\right) \right] - \frac{b^{2}\xi}{2} (2\xi + 1) \frac{1}{\cosh^{2}\left(\frac{b}{2}(x-x_{0})\right)}.
\end{align}
These potentials belong to the category of hyperbolic Rosen-Morse potentials and satisfy the shape invariance condition.  
Here, \( x_0 \) is defined as $ x_0 \equiv -\frac{1}{b} \ln a$.  
It can be explicitly shown that the partner potentials in \eqref{RMpot} satisfy the shape invariance relation  
\begin{align} \label{shapeinv}
V^{(2)}(x, \alpha_1) = V^{(1)}(x, \alpha_2) + R(\alpha_2),
\end{align}  
where \( \alpha_2 \) is a function of \( \alpha_1 \), i.e., \( \alpha_2 = \alpha_2 (\alpha_1) \). This implies that the potential \( V^{(1)}(x, \alpha_2) \) transforms into \( V^{(2)}(x, \alpha_1) \) by changing the parameter from \( \alpha_1 \) to \( \alpha_2 \) and adding a constant term \( R(\alpha_2) \).

More comments are given on the shape invariant property of the Rosen-Morse potentials in \eqref{RMpot}.  
General forms of the Rosen-Morse potential and its superpartner are given by~\cite{Dutt:1986va}  
\begin{align}\label{RMpot2}
V^{(1)}(x;p, q, r)  &= p^2 + \frac{q^2}{p^2} + 2 q \tanh (r x) -\frac{p (p + r)}{\cosh^2 (r x)}, 
\nonumber\\
V^{(2)}(x;p, q, r)  &=  p^2 + \frac{q^2}{p^2} + 2 q \tanh (r x) -\frac{p (p - r)}{\cosh^2 (r x)},
\end{align}
where \( p \), \( q \), and \( r \) are constant parameters.  
The shape invariant property in \eqref{shapeinv} for the general forms in \eqref{RMpot2} is given by  
\begin{align}\label{shapeinv2}
V^{(2)}(x;p,q,r) = V^{(1)}(x;p-r, q, r) + R(p-r, q,r),
\end{align}
where  
\begin{align}
R(p-r, q,r) = p^2 - \big(p- r  \big)^2 + q^2 \Bigg(\frac{1}{p^2} - \frac{1}{(p-r)^2}\Bigg).
\end{align}
The potentials in \eqref{RMpot} are obtained from those in \eqref{RMpot2} by setting  
\begin{align}
p = - b \xi, \quad q = b^2 \xi^2, \quad r= \frac{b}{2}.
\end{align}
Then, the relation in \eqref{shapeinv2} is also satisfied with this setting.  

Using this shape invariance property of the two potentials, one can determine the spectrum and generalized eigenstates of the quantum mechanical system. We apply this method to find mode solutions of the scalar field \( \phi \).

\subsection{Mode solutions and canonical quantization}
\label{subsec2.3}
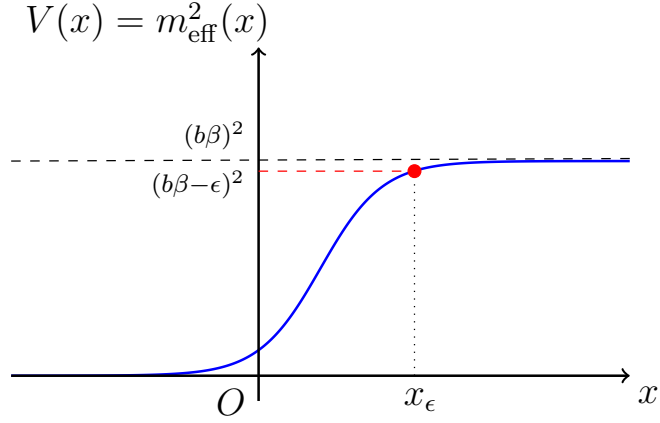
\begin{figure}
\begin{center}
\begin{tikzpicture}[scale=1.2]
\begin{axis}[
        axis lines =  center ,
        axis y line=none,
        axis x line=none,
        xlabel = $x$,
        ylabel = {$V(x)=m^2_{\rm eff}(x)$},
        xmin=-5, xmax=5,
        ymin=-0.3, ymax=2.6,
        xtick=\empty,
        ytick=\empty,
        xlabel style={below right},
        ylabel style={left above},
        clip=false,
        y=1.4cm
     ]
       
    \addplot[thick, blue, smooth, domain=-5:5, samples=200] {0.85*(tanh(x) + 1)};
    \draw[->,thick] (axis cs:-1,-0.2) -- (axis cs:-1,2.6)  ;
    \node at (axis cs:-2.8,2.8)  {${\small V(x)=m^2_{\rm eff}(x)}$};
     \draw[->, thick] (axis cs:-5,0) -- (axis cs:5,0)  ;
     \node at (axis cs:5.3,-0.15)  {$x$};

    \node[below left] at (axis cs:-1,0) {$O$};
    \node[left] at (axis cs:-1,1.9) {${\scriptstyle (b\beta)^2}$};
    \draw[dashed] (axis cs:-5,1.7) -- (axis cs:5,1.72);
    
    \addplot[mark=*, only marks, red] coordinates {(1.52, 1.62)};
    
    \draw[red, dashed] (axis cs:-1,1.62) -- (axis cs:1.7,1.62);
    
    \draw[dotted] (axis cs:1.52,0) -- (axis cs:1.52,1.62);
    \node[above right] at (axis cs:-2.95 ,1.3) {${\scriptstyle (b \beta - \epsilon)^2}$};
    \node[below] at (axis cs:1.62,0) {$x_\epsilon$};
    
 %
\end{axis}
\end{tikzpicture}
\caption{ We depict the  monotonically increasing, S-shaped graph of  \( m_{\text{eff}}^2(x) \) as a function of \( x \). Here, \( x_\epsilon \) denotes the position of an observer located slightly out of the right asymptotic infinity. When we compute the positive frequency Wightman function $G^+$, we disregard the modes  \( \Phi_\epsilon \) within the energy range between \( b \beta - \epsilon \) and \( b \beta \), where \( \frac{\epsilon}{b \beta} \ll 1 \).
}\label{Fig.1}
\end{center}
\end{figure}

In the previous subsection, we obtained two partner potentials in \eqref{RMpot} for $a > 0$, where the geometry (or the mass function $m_{{\rm eff}}^2(x)$) is defined over the entire spatial region. 
We now further restrict the parameter range to $\xi \ge \frac{1}{4}$, where $V^{(1)}(x) = m_{{\rm eff}}^2(x)$ is a monotonically increasing function that approaches zero as $x \to -\infty$ and $(m_0/a)^2$ as $x \to +\infty$. See Fig.~\ref{Fig.1}. 
This particular form of the position-dependent mass function has various physical applications, such as the expansion of a Higgs condensate bubble in the early universe~\cite{Coleman:1977py,Callan:1977pt,Linde:1981zj,Moore:1995si,Bodeker:2009qy, Bodeker:2017cim, Kubota:2024wgx}.

Using the technique of SQM for the shape-invariant potential in \eqref{RMpot}, one can express the mode solutions of the equation \eqref{Aphiw} in terms of hypergeometric functions as
\begin{align}\label{phiwL}
\phi_{\omega}(y) = (1+y)^{\beta} \left[ a_{1} y^{\alpha} F(A,B; C \,|\, -y) + a_{2} \, y^{\alpha + 1 - C} F(A - C + 1, B - C + 1; 2 - C \,|\, -y) \right],
\end{align}
where $a_{1}$ and $a_{2}$ are integration constants, $y \equiv a e^{bx} = e^{b(x - x_0)}$, $\alpha = \frac{i \omega}{b}$, $\beta = 2\xi = \frac{m_0}{ab}$, and
\begin{align}\label{ABC}
A = \frac{i}{b}\left(\omega - k \right) + \beta, \qquad B = \frac{i}{b}\left(\omega + k\right) + \beta, \qquad C = 1 + \frac{2 i \omega}{b},
\end{align} 
with $k^2 \equiv \omega^2 - (2 b \xi)^2$.

In the left asymptotic region $x \to -\infty$ ($y \to 0$), the mode solutions in \eqref{phiwL} are expressed as
\begin{align}\label{fwxminf}
\phi_\omega(x) \underset{x \to - \infty}{\longrightarrow} a_{1} \, e^{i\omega (x - x_{0})} + a_{2} \, e^{-i\omega (x - x_{0})}.
\end{align}
Therefore, the mode solutions in \eqref{phiwL} are suitable for physical applications, such as canonical quantization, around the left asymptotic region.

On the other hand, by using the linear transformation of the hypergeometric function,
\begin{align}
\frac{\sin \pi(B - A)}{\pi \Gamma(C)} F(A, B; C \,|\, z) &= \frac{(-z)^{-A}}{\Gamma(B)\Gamma(C - A)\Gamma(A - B + 1)} F\Big(A, A - C + 1; A - B + 1 \,\Big|\, \frac{1}{z} \Big) \nonumber \\
&  - \frac{(-z)^{-B}}{\Gamma(A)\Gamma(C - B)\Gamma(B - A + 1)} F\Big(B, B - C + 1; B - A + 1 \,\Big|\, \frac{1}{z} \Big),
\end{align}
we can rewrite the mode solutions \eqref{phiwL} as
\begin{align}\label{phiwR}
\phi_{\omega}(y) = (1 + y)^{\beta} \Big[ b_{1} \, y^{\alpha - A} F\Big(A, A - C + 1; A - B + 1 \, \Big|\, -\frac{1}{y} \Big)  \nonumber \\
 + b_{2} \, y^{\alpha - B} F\Big(B, B - C + 1; B - A + 1 \, \Big|\, -\frac{1}{y} \Big) \Big],
\end{align}
where the constants $b_{1}$ and $b_2$ are related to the constants $a_{1}$ and $a_{2}$ from \eqref{phiwL} as follows:\footnote{The reflection coefficient $R_{\omega} \in \mathbb{C}$ is defined by
\begin{equation} \nonumber
R_{\omega} \equiv -\frac{\Gamma(C)\Gamma(A - C + 1)\Gamma(1 - B)}{\Gamma(2 - C)\Gamma(A)\Gamma(C - B)} = -\frac{\Gamma\left(1 + 2i\frac{\omega}{b}\right) \Gamma\left(\beta - \frac{i}{b}(\omega + k)\right) \Gamma\left(1 - \beta - \frac{i}{b}(\omega + k)\right)}{\Gamma\left(1 - 2i\frac{\omega}{b}\right) \Gamma\left(\beta + \frac{i}{b}(\omega - k)\right) \Gamma\left(1 - \beta + \frac{i}{b}(\omega - k)\right)}
\end{equation}
with the relation $R^{*}_{\omega} = \frac{1}{R_{-\omega}}$.}
\begin{align}
b_{1} &= \frac{\Gamma\left(2i\frac{k}{b}\right)\Gamma\left(1 + 2i\frac{\omega}{b}\right) [a_{1} - R^{*}_{\omega} \, a_{2}]}{\Gamma\left(\beta + \frac{i}{b}(\omega + k) \right) \Gamma\left(1 - \beta + \frac{i}{b}(\omega + k) \right)}, \nonumber \\
b_{2} &= \frac{\Gamma\left(-2i\frac{k}{b}\right)\Gamma\left(1 - 2i\frac{\omega}{b}\right) [-R_{\omega} a_{1} + a_{2}]}{\Gamma\left(\beta - \frac{i}{b}(\omega + k)\right) \Gamma\left(1 - \beta - \frac{i}{b}(\omega + k)\right)}.
\end{align} 
In the right asymptotic region $x \to +\infty$ ($y \to \infty$), the leading asymptotic behavior of the mode solutions \eqref{phiwR} is given by
\begin{align}
\phi_\omega(x) \underset{x \to + \infty}{\longrightarrow} b_{1} \, e^{ik (x - x_{0})} + b_{2} \, e^{-ik (x - x_{0})}.
\end{align}
Thus, the mode solutions \eqref{phiwR} are useful for canonical quantization around the right asymptotic region.

We now quantize the scalar field $\phi$ using the mode solutions \eqref{phiwL} and \eqref{phiwR}. First, we expand the scalar field using the mode solutions in \eqref{phiwL}  schematically   as follows:
\begin{align}\label{phiL}
    \phi_{\rm L}(\pmb{x}) &= \int_{0}^{\infty}\frac{d\omega}{\sqrt{2\pi}} \frac{1}{\sqrt{2\omega}} \sum_{i=\pm}
    \left[ a_{\omega}^{(i)} u^{(i)}_{\omega}(\pmb{x}) + \big(a_{\omega}^{(i)}\big)^{\dagger} \big(u^{(i)}_{\omega}(\pmb{x})\big)^{*} \right], 
\end{align}
where $u^{(\mp)}_{\omega}(\pmb{x})$ are given by\footnote{\label{abouta}  We define \( a \equiv e^{-b x_0} \) and perform a change of coordinates \( x - x_0 \mapsto x \). In this new coordinate system, the Ricci scalar in \eqref{R} is given by
\(
\mathcal{R} = 2 a^2 b^2 e^{-bx}.
\)
}
\begin{align}
    u^{(-)}_{\omega}(\pmb{x}) &= (1 + e^{bx})^{\beta} F(A, B; C \,|\, - e^{bx}) e^{-i\omega (t - x)}, \label{um} \\
    u^{(+)}_{\omega}(\pmb{x}) &= (1 + e^{bx})^{\beta} F(A - C + 1, B - C + 1; 2 - C \,|\, - e^{bx}) e^{-i\omega (t + x)}, \label{up}
\end{align}
and the complex conjugates of these mode functions are given by
\begin{align}
    \big(u^{(-)}_{\omega}(\pmb{x})\big)^{*} &= (1 + e^{bx})^{2\xi} F(A - C + 1, B - C + 1; 2 - C \,|\, - e^{bx}) e^{i\omega (t - x)}, \label{umcc} \\
    \big(u^{(+)}_{\omega}(\pmb{x})\big)^{*} &= (1 + e^{bx})^{2\xi} F(A, B; C \,|\, - e^{bx}) e^{i\omega (t + x)}. \label{upcc}
\end{align}
Note that
\[
\big[F(A, B; C \,|\, - e^{bx})\big]^{*} = F(A - C + 1, B - C + 1; 2 - C \,|\, - e^{bx}),
\]
as seen from \eqref{ABC}.

In principle, the normalization is determined through the Klein-Gordon inner product. 
If the mode functions \eqref{um} and \eqref{up} were correctly orthonormalized, our normalization convention for $u^{(\mp)}_{\omega}(\pmb{x})$ would be consistent with the non-vanishing commutator of the creation and annihilation operators, expressed as
\begin{equation} \label{aCom}
    [a_{\omega}^{(i)}, (a_{\omega'}^{(j)})^{\dagger}] = \delta^{ij} \delta(\omega - \omega'),
\end{equation}
where $i$ and $j$ denote $-$ or $+$.

However, the operators \( a^{(-)}_{\omega} \) and \( a^{(+)}_{\omega} \) are not independent but proportional in the range \( 0 \le \omega <  b\beta \), due to the absence of propagating degrees of freedom in the limit \( x \to +\infty \). Consequently, the commutation relation given in \eqref{aCom} is valid only for \( \omega \gg b\beta \).  
Mode expansion and quantization become involved in the low-frequency regime, resulting from reflecting modes from the mass function wall within the range $ 0 \leq \omega < b \beta $.  To address this issue, a proposal has been made in the context of the spectral method~\cite{Kubota:2024wgx}, in which the mode functions can be properly normalized without relying on the Klein-Gordon inner product. According to the spectral method, the field expansion \eqref{phiL} should be rewritten in terms of new operators and new mode functions constructed by\footnote{In fact, this construction is indirect and uses a mathematical tool known as a Weyl-Titchmarsh-Kodaira theory~\cite{Kubota:2024wgx}.}  linear combinations of \( u^{(i)}_{\omega}(\pmb{x}) \) and \( (u^{(i)}_{\omega}(\pmb{x}))^{*} \). 
Before going ahead, let us elaborate on the asymptotic behavior of the schematic expression in~\eqref{phiL}. In the left asymptotic region $(x \to -\infty)$, the mode functions $u^{(\mp)}_{\omega}(\pmb{x})$ reduce to
\begin{align}
    u^{(\mp)}_{\omega}(\pmb{x}) \underset{x \to -\infty}{\longrightarrow} e^{-i\omega(t \mp x)},
\end{align}
and the field operator in equation \eqref{phiL} approaches
\begin{align}    \label{phiLnew}
  \phi_{\rm L}(\pmb{x}) & \underset{x \to -\infty}{\simeq} \int^{\infty}_{b\beta} \frac{d\omega}{\sqrt{2\pi}} \frac{1}{\sqrt{2\omega}} \Big[ a_{\omega}^{(+)} e^{-i\omega (t + x)} + a_{\omega}^{(-)} e^{-i\omega (t - x)} + h.c.~ \Big]  \nonumber  \\
& \qquad \quad+ \int^{b\beta}_{0} \frac{d\omega}{\sqrt{2\pi}} \frac{1}{\sqrt{2\omega}}   \Big[ c_{\omega}\, e^{-i\omega (t + x)} + r_{\omega}c_{\omega}\, e^{-i\omega (t - x)} +h.c.  ~\Big] \,,
\end{align}
where $c_{\omega}$ is a new operator  alluded above, $r_{\omega}$ denotes a certain phase factor, and  $h.c.$ means the Hermitian conjugate. This expression reveals that the quantization describes the system in terms of massless particles in the $x \to - \infty$ region. As highlighted earlier, this field quantization deviates from conventional canonical quantization due to the interdependence of \( u^{(-)}_{\omega} \) and \( u^{(+)}_{\omega} \) in the low-frequency regime. 
 We have coined this quantization scheme ``L-quantization'' in~\cite{Ho:2022omx}.
Within the L-quantization framework, the vacuum state \( |0\rangle_{\rm L} \) is defined as the state annihilated by \( a_{\omega}^{(\mp)} \) and $c_{\omega}$:
\begin{equation} \label{leftvacuum}
    a_{\omega}^{(\mp)} |0\rangle_{\rm L} = c_{\omega}|0\rangle_{\rm L}= 0\,.
\end{equation}
In this study, our analysis focuses solely on the right asymptotic region. We defer a detailed investigation of L-quantization in the left asymptotic region to future research.

 While the above  L-quantization would be appropriate for the left asymptotic region, it does not extend to the right asymptotic region, rendering the quantization scheme unsuitable for that domain or for an observer situated there. 
We now introduce another quantization scheme, referred to as ``R-quantization.'' From \eqref{phiwR}, we define $v^{(\mp)}_{k}(\pmb{x})$ as
\begin{align}
    v^{(-)}_{k}(\pmb{x}) &= (1 + e^{-bx})^{2\xi} F\left(A, A - C + 1; A - B + 1 \, \Big|\, -e^{-bx} \right) e^{-i (\omega t - kx)}, \label{vm} \\
    v^{(+)}_{k}(\pmb{x}) &= (1 + e^{-bx})^{2\xi} F\left(B, B - C + 1; B - A + 1 \, \Big|\, - e^{-bx} \right) e^{-i (\omega t + kx)}, \label{vp}
\end{align}
where $\omega = \sqrt{k^{2} + b^{2}\beta^{2}}$. Since the relation $\big[F(A, A - C + 1; A - B + 1 \,|\, - e^{-bx})\big]^{*} = F(B, B - C + 1; B - A + 1 \,|\, - e^{-bx})$ holds, the complex conjugates of the mode functions are given by
\begin{align}
    \big(v^{(-)}_{k}(\pmb{x})\big)^{*} &= (1 + e^{-bx})^{2\xi} F(B, B - C + 1; B - A + 1 \,|\, - e^{-bx}) e^{i (\omega t - kx)}, \label{vmcc} \\
    \big(v^{(+)}_{k}(\pmb{x})\big)^{*} &= (1 + e^{-bx})^{2\xi} F(A, A - C + 1; A - B + 1 \,|\, - e^{-bx}) e^{i (\omega t + kx)}. \label{vpcc}
\end{align}
In the right asymptotic region, $v^{(\mp)}_{k}(\pmb{x})$ reduce to
\begin{align}
    v^{(\mp)}_{k}(\pmb{x}) \underset{x \to +\infty}{\longrightarrow} e^{-i(\omega t \mp kx)},
\end{align}
and the canonically quantized form of the scalar field in the right asymptotic region $(x \to +\infty)$ reduces to
\begin{align}
    \phi_{\rm R(\rm M)}(\pmb{x}) \equiv \phi_{\rm R}(\pmb{x})|_{x \to +\infty} &= \int^{\infty}_{0} \frac{dk}{\sqrt{2\pi}} \frac{1}{\sqrt{2\omega}} \Big[ b_{k}^{(+)} e^{-i (\omega t + kx)} + b_{k}^{(-)} e^{-i (\omega t - kx)} \nonumber\\
    &\qquad + \big(b_{k}^{(+)}\big)^{\dagger} e^{i (\omega t + kx)} + \big(b_{k}^{(-)}\big)^{\dagger} e^{i (\omega t - kx)} \Big].
\end{align}
Similar to \eqref{aCom}, the non-vanishing commutator of the creation and annihilation operators is given by
\begin{equation} \label{bCom}
    [b_{k}^{(i)}, (b_{k'}^{(j)})^{\dagger}] = \delta^{ij} \delta(k - k').
\end{equation}
Just as in L-quantization, we define the vacuum in R-quantization, $|0\rangle_{\rm R(\rm M)}$, as
\begin{equation} \label{rightvacuum}
    b_{k}^{(\mp)} |0\rangle_{\rm R(\rm M)} = 0,
\end{equation}
where $|0\rangle_{\rm R(\rm M)}$ is identical to the Minkowski vacuum $|0\rangle_{\rm M}$. The above equation shows that, in contrast to L-quantization, R-quantization leads to particles with a mass of $b\beta = \frac{m_0}{a} = 2\xi b$, as inferred from $\omega = \sqrt{k^2 + b^2\beta^2}$. This implies that (local) `right' observers cannot detect massless particles with energy $\omega < b\beta$, which can be observed in the `left' region.

Now, we consider the field operator slightly out of the right asymptotic region. 
Proper normalization of the mode functions defined in  \eqref{vm} and \eqref{vp} via the Klein-Gordon inner product is generally difficult to achieve,   as L-quantization in the left region.  As an alternative, the spectral method outlined earlier can be employed to obtain a properly normalized quantization, yielding results consistent with the L-quantization scheme. Nevertheless, as demonstrated below, within the range of approximations adopted in this study, the R-quantization based solely on the mode functions $v^{(i)}_k$ and $(v^{(i)}_k)^\dagger$ provides consistent and reliable outcomes.
Specifically, for the R-quantization slightly out of the right asymptotic region, the field operator can be expressed in terms of the mode functions introduced in \eqref{vm} and \eqref{vp} as follows: 
\begin{align} \label{phiR}
    \phi_{\rm R}(\pmb{x}) & =  \int_{0}^{\infty}\frac{dk}{\sqrt{2\pi}}\frac{1}{\sqrt{2\omega}} \sum_{i=\pm}
    \left[ b_{k}^{(i)} v^{(i)}_{k}(\pmb{x}) + \big(b_{k}^{(i)}\big)^{\dagger}\big(v^{(i)}_{k}(\pmb{x})\big)^{*} \right] + \Phi_\epsilon (\pmb{x}),
\end{align}
  where  $\Phi_\epsilon (\pmb{x})$ accounts for the  exponentially decaying  modes  as $x\to\infty$ in the energy range $0 \leq \omega < b \beta$.  In this paper, the main focus is physics at the position $x=x_{\epsilon}$ slightly out of the right asymptotic region (See~Fig.\ref{Fig.1}).  One may worry about the contribution of propagating modes near the position $x=x_{\epsilon}$ over the range $b\beta -\epsilon \le \omega < b\beta$. To give an order estimate near $x=x_{\epsilon}$, let us write~\eqref{phiLnew} with $W(\pmb{x}) \sim e^{- i(\omega t \pm kx)}$  schematically as  
\begin{align} \label{Phieps}
\Phi_\epsilon(\pmb{x}) = \int_{b \beta - \epsilon}^{b \beta}\frac{d\omega}{\sqrt{2\pi}} \frac{1}{\sqrt{2\omega}} \left[ c_\omega W_{\omega}(\pmb{x}) + c_\omega^\dagger (W_{\omega}(\pmb{x}))^* \right]  + {\cal O}(e^{-bx_{\epsilon}})\,.
\end{align}
where the exponentially decaying part comes from the range $0\le \omega < b\beta -\epsilon$. As discussed in Appendix~\ref{AppB}, the  first integral term over the range $b\beta -\epsilon < \omega < b\beta$ contributes  at a higher order. Then, we can safely neglect the contribution of $\Phi_{\epsilon}$, which is ${\cal O}(e^{-2bx_{\epsilon}})$,  to  2-point function $\langle \phi \phi \rangle $,  since our analysis  of  the 2-point function $\langle \phi \phi \rangle $ focuses on terms up to  ${\cal O}(e^{-bx_{\epsilon}})$.

By adopting the approximation $\frac{\epsilon}{b \beta} \ll 1$ and omitting \( \Phi_\epsilon(x) \), the field operator is simplified, though this comes at the expense of completeness. This approximation neglects certain field contributions, potentially affecting the canonical commutation relations. Consequently, the simplified field operator may not fully satisfy these relations, but we can take
\begin{align}
[b_{k}^{(i)}, (b_{k'}^{(j)})^{\dagger}] \simeq \delta^{ij} \delta(k - k').
\end{align}
Within this framework, we introduce the vacuum $|0\rangle_{{\rm R}}$, defined by
\begin{align}\label{bkR2}
b_{k}^{(\mp)} |0\rangle_{\rm R} = 0.
\end{align}
Since the term $\Phi_\epsilon(\pmb{x})$ in \eqref{phiR} is omitted under this approximation, an ambiguity remains in fully defining $|0\rangle_{{\rm R}}$. Nevertheless, 
 it turns out that the above approximation scheme is sufficient to extract meaningful physical effects, as will be shown in the following sections.

Up to this point, we have discussed the canonical quantization of the scalar field from the perspective of IFT, which can also be understood as quantization within the FTCS framework for the right asymptotic region. This region corresponds to the weak curvature limit, where the FTCS framework is suitable for our analysis. Thus,  R-quantization is valid within both the FTCS and IFT frameworks. However, L-quantization is not directly applicable to the FTCS framework due to the presence of a naked singularity as $x \to -\infty$. We leave the issue of the quantization and its implications near the right asymptotic region for future investigation~\cite{HKPY}.

\section{Hadamard Two Point function}\label{sec3}

In this section, we aim to calculate the two-point correlation function in the right asymptotic region of the background given by \eqref{ourmet}, where the FTCS description is valid. In section \ref{sec5}, we will reinterpret the FTCS results within the context of IFT.

\subsection{Quantum states in our background spacetime}\label{subsec3.1}

One of the fundamental problems in QFTCS is the ambiguity in defining an appropriate vacuum state for field quantization. This difficulty arises from the absence of a unique, globally defined notion of positive frequency modes in a generic curved spacetime, which are essential for defining particles and the vacuum state.
  To address such a conceptual problem, we adopt the algebraic formulation of QFTCS,  incorporating several concepts studied in the research area of algebraic QFT~\cite{Haag:1992hx,Wald:1995yp,Fredenhagen:2014lda}.

In general, this approach is typically framed in the context of a global Hadamard state. However, due to the existence of a null curvature singularity in the limit as $x \to -\infty$ in our case~\cite{Ho:2022omx}, it appears that the Hadamard condition cannot be globally satisfied. To address this issue, we will consider a global algebraic state that satisfies the Hadamard condition locally. Accordingly, we will extend the concept of Hadamard renormalization for our two-point function calculation\footnote{For related discussions, see \cite{Kwon:2022fhv}.}.

The state $|0\rangle_{\rm R(M)}$ introduced in \eqref{rightvacuum} was identified as the appropriate vacuum for the right asymptotic region. Subsequently, the R-vacuum $|0\rangle_{\rm R}$ was also considered for a region slightly out of the right asymptotic region \eqref{bkR2}. While modes with energy in the range $b \beta - \epsilon \le \omega < b \beta$ $(0< \frac{\epsilon}{b \beta} \ll 1)$ exist in this region (see Figure \ref{Fig.1}), the R-vacuum $|0\rangle_{\rm R}$ cannot accommodate these modes. In other words, particles within the energy range $b \beta - \epsilon \le \omega < b \beta$ cannot be included in the Fock space $\mathcal{F}_{{\rm R}}$ constructed from the R-vacuum.

To construct a conceptually satisfactory framework, we introduce an algebraic state, $\omega_{\rm R}^{\epsilon}$\footnote{$|0\rangle_{\rm R}^\epsilon$ is represented by $\omega_{\rm R}^{\epsilon}$.}, which is defined as a positive, normalized linear functional that assigns complex numbers to field operators. The two-point correlation function with respect to $|0\rangle_{\rm R}^{\epsilon}$ is then identified with the complex value assigned by $\omega_{\rm R}^{\epsilon} (\hat{\phi} \hat{\phi})$. Here, $\epsilon$ refers to a specific local region $x_\epsilon$, defined with $\epsilon$ in Figure \ref{Fig.1}. Using the Gelfand-Naimark-Segal (GNS) construction, we can return to the conventional Hilbert space representation with the algebraic state $\omega_{\rm R}^{\epsilon}$. This construction produces a representation of the algebra on a Hilbert space, $\mathcal{H}_{\omega_{\rm R}^{\epsilon}}$, where $|0\rangle_{\rm R}^{\epsilon}$ serves as the vacuum state of $\mathcal{H}_{\omega_{\rm R}^{\epsilon}}$ in our context.
If the Hilbert space $ \mathcal{H}_{\omega_{\rm R}^{\epsilon}} $ were to provide a consistent particle interpretation across the entire system, it would need to include states generated by scattering processes, such as transitions between massless and massive particles. However, as discussed in \cite{Ho:2022omx}, it is strongly anticipated that, in our model governed by the quadratic Lagrangian \eqref{SIFTf}, the Hilbert space $\mathcal{H}_{\omega_{\rm R}^{\epsilon}}$ consists of quasi-free states, which are unable to account for scattering phenomena. 
Furthermore, this is inconsistent with the fact that only massive particles exist in the right asymptotic region, whereas only massless particles exist in the left asymptotic region.

\subsection{Hadamard regularization of the two-point function}\label{subsec3.2}

In this subsection, we obtain a regularized two-point function in QFTCS by adapting the Hadamard regularization method.
To begin, we calculate the VEV of the product of two field operators at a position at $x_\epsilon$ slightly displaced from the right asymptotic region, using the aforementioned vacuum \( \omega_{\mathrm{R}}^\epsilon \).
We approximate this VEV as
\begin{align}\label{approx}
\omega_{\mathrm{R}}^\epsilon (\phi({\pmb x}) \phi({\pmb x}')
 ) \simeq {}_{{\rm R}}^{\,\epsilon} \!\langle 0|\phi({\pmb x}) \phi({\pmb x}')|0\rangle_{{\rm R}}^\epsilon.
\end{align}

Now, let us consider the Hadamard regularization of the two-point function. 
Specifically, we focus on the (positive frequency) Wightman function, $ G^{+}({\bf x}, {\bf x}') $, which is defined by 
\begin{align}
(\Box - m_0^2  - \xi \mathcal{R} ) G^{+} ({\pmb x}, {\pmb x}')=- \frac{1}{\sqrt{-g}} \delta(t-t') \delta(x-x'), 
\end{align}
subject to appropriate boundary conditions.
In an abstract form, the Wightman function can be written in terms of mode function given in \eqref{uw},
\begin{equation} \label{}
G^{+}({\pmb x}, {\pmb x}') = \int d\mu_{\eta}\,   u_{\eta}({\pmb x}) u^{*}_{\eta}({\pmb x}') \,,
\end{equation}
where $\mu_{\eta}$ denotes the measure index of mode function spaces.

In our approximation, the Wightman function can be identified as
\begin{align}
    G_\epsilon^{+} ({\pmb x}, {\pmb x}') \simeq {} {}_{{\rm R}}^{\,\epsilon} \!\langle 0|\phi({\pmb x}) \phi({\pmb x}')|0\rangle_{{\rm R}}^\epsilon,
\end{align}
which exhibits a specific singular structure in the coincidence limit as \( {\pmb x}' \) approaches \( {\pmb x} \). In two-dimensional spacetime, the singular structure of the Wightman function is well-known and can be expressed as
\begin{align}\label{G+xxp}
    G^{+} ({\pmb x}, {\pmb x}')
    = \frac{1}{4 \pi} \left( V({\pmb x}, {\pmb x}') \ln \left[\mu^2 \sigma({\pmb x}, {\pmb x}')\right] + W({\pmb x}, {\pmb x}';\mu) \right),
\end{align}
where \( V({\pmb x}, {\pmb x}') \) and \( W({\pmb x}, {\pmb x}';\mu) \) are symmetric biscalar functions that remain regular as \( {\pmb x}' \) approaches \( {\pmb x} \). The biscalar function \( V({\pmb x}, {\pmb x}') \) is  determined by  the geometry and the field equation. On the other hand, the biscalar function \( W({\pmb x}, {\pmb x}';\mu) \) encodes quantum state dependence in addition to the geometry and the field equation. 
The parameter \(\mu\) is an arbitrary constant with mass dimension one, chosen such that the vacuum energy in Minkowski space is set to zero. 
The biscalar function \(2\sigma(\pmb{x}, \pmb{x}')\), known as the Synge function, is defined as the square of the geodesic distance between the points \(\pmb{x}\) and \(\pmb{x}'\).

The Wightman function can be calculated with  the commutation relation in \eqref{bCom} 
\begin{align}    \label{G+xx}
G_\epsilon^{+} ({\pmb x}, {\pmb x}') \simeq {} \int^{\infty}_{0} \frac{dk}{4 \pi \omega_k} \sum_{i=\mp} v^{(i)}_{k}({\pmb x}) (v^{(i)}_{k}({\pmb x'}))^{*}, 
\end{align}
where the mode functions,  $v^{(\mp)}_{k}({\bf x})$ and  $(v^{(\mp)}_{k}({\pmb x'}))^{*}$, are given in \eqref{vm} - \eqref{vpcc} and $\omega_k \equiv  \sqrt{k^{2}+(b\beta)^2}$. Here, the approximation in \eqref{G+xx} represents that we have omitted the terms related with $\Phi_\epsilon(\pmb{x})$ in \eqref{phiR}. In the following, we will focus on the Wightman function of spacelike separated points ${\pmb x}$ and ${\pmb x}'$, for simplicity. 
Although we use a timelike point-splitting, the VEV of the energy-momentum tensor remains unchanged after applying the Hadamard renormalization prescription for a spacelike point-splitting.

Before going ahead let us recall that the 
$(1+1)$-dimensional positive frequency Wightman function 
for a massive scalar field of mass $m_0$ on Minkowski spacetime $G^{+}_{{\rm M}}$ is given by
\begin{align}   \label{GM+}
G^{+}_{{\rm M}}({\pmb x}, {\pmb x}')  =  \frac{1}{2\pi}
 K_{0}\Big(m_0|{\pmb x}-{\pmb x}'|\Big)\,,
\end{align}
where $|{\pmb x}-{\pmb x}'|$ means $\sqrt{-(t-t'-i\varepsilon)^{2} + (\vec{x} - \vec{x}')^{2}}$, $K_0$ is the modified Bessel function of the second kind, and $\varepsilon$ is inserted  for the positive frequency Wightman function. In the coincident limit, the Wightman function in \eqref{GM+} is expanded by \begin{equation} \label{}
G^{+}_{{\rm M}}({\pmb x}, {\pmb x}') \sim - \frac{1}{2\pi}\ln\left(m_0 |{\pmb x}-{\pmb x}'|\right) +\frac{1}{2 \pi} \left (- \gamma + \ln 2 \right) 
  +\cdots\,,
\end{equation}
where $\gamma=0.5772...$ is the Euler constant.

Now turn to our case. 
As mentioned above, we use the Hadamard regularization method to renormalize the Wightman function \(G_\epsilon^+(\pmb{x}, \pmb{x}')\) in \eqref{G+xxp}. The renormalized Wightman function is then given by
\begin{align}\label{regpart0}
G^{+}_{\text{ren}} (\pmb{x}, \pmb{x}'; \mu) =  \frac{1}{4 \pi}  W (\pmb{x}, \pmb{x}';\mu)
= G_\epsilon^{+} (\pmb{x}, \pmb{x}')  
- \frac{1}{4 \pi} V(\pmb{x}, \pmb{x}') \ln  \left[ \mu^2\sigma (\pmb{x}, \pmb{x}') \right].
\end{align}
To obtain a finite VEV of the energy-momentum tensor at a position $x_\epsilon$ slightly displaced from asymptotic right infinity, we need the renormalized Wightman function \(G^{+}_{\text{ren}} (\pmb{x}, \pmb{x}';\mu)\) as \(\pmb{x}' \to \pmb{x}\). For this purpose, we must expand \(\sigma (\pmb{x}, \pmb{x}')\), \(V(\pmb{x}, \pmb{x}')\), and  \(G_\epsilon^{+} (\pmb{x}, \pmb{x}')\) in the right-hand side of \eqref{regpart0}.

At first, the expanded form of the Synge function\footnote{See the footnote \ref{abouta} for the parameter $a$. 
} in the background given by \eqref{ourmet} is
\begin{align} \label{Synexp}
a^2 \,\sigma (\pmb{x}, \pmb{x}') = & \frac{1}{2} \left[ -(t - t')^2 + (x - x')^2 \right]  
+ e^{-b \frac{x + x'}{2}} \Big[
  (t - t')^2 - (x - x')^2 
  \nonumber\\
& \quad - \frac{b^2}{24} (x - x')^4 + \frac{b^2}{24} (t - t')^2 (x - x')^2 + \cdots \Big]
+ \mathcal{O}(e^{-b(x+ x')}),
\end{align}
where the ellipsis \(\cdots\) within the second bracket represents higher-order terms involving combinations of \((t - t')\) and \((x - x')\).
The expansion in \eqref{Synexp} is consistent with various known properties of the Synge function.

We now summarize several properties of the Synge function \(\sigma(\pmb{x}, \pmb{x}')\). It satisfies the following equations:
\begin{align}    \label{syngediffeq1}
2 \sigma &=  \nabla^{\mu} \sigma  \nabla_{\mu} \sigma,
\\
\label{syngediffeq2}
\Box_{\pmb{x}} \sigma &= 2 - 2  \Delta ^{-1/2} \nabla_\mu \Delta ^{1/2} \nabla^\mu\sigma,
\end{align}
where \(\Delta  (\pmb{x}, \pmb{x}')\) is the biscalar form of the van Vleck-Morette determinant:
\begin{align}    \label{}
\Delta  (\pmb{x}, \pmb{x}') =  - [-g(\pmb{x})]^{-1/2}  \det ( -\nabla_{\nu'}\nabla_\mu \sigma(\pmb{x}, \pmb{x}')) [-g(\pmb{x}')]^{-1/2} 
\end{align}
with the boundary condition
\begin{align}    \label{Dxxp}
 \lim_{\pmb{x'} \to \pmb{x}} \Delta  (\pmb{x}, \pmb{x}') =  1.
\end{align}
An immediate consequence of the above equations \eqref{syngediffeq1} and \eqref{syngediffeq2} is
\begin{align}    \label{expandingdiffsynge}
\nabla_\mu\nabla_\nu \sigma = g_{\mu \nu} - \frac{1}{3} \mathcal{R}_{\mu \alpha \nu \beta} \nabla^\alpha \sigma\nabla^\beta \sigma  + \mathcal{O} (\sigma^{3/2}).
\end{align}

Now, we turn to the biscalar functions \(V(\pmb{x}, \pmb{x}')\) and  \(W(\pmb{x}, \pmb{x}';\mu)\) in \eqref{regpart}, which can be expanded in terms of the Synge function \(\sigma (\pmb{x}, \pmb{x}')\), 
\begin{align}\label{VWnxxp}
&V(\pmb{x}, \pmb{x}') = \sum_{n=0}^{+\infty} V_n(\pmb{x}, \pmb{x}') \sigma^n(\pmb{x}, \pmb{x}'), 
\nonumber\\
&W(\pmb{x}, \pmb{x}';\mu) = \sum_{n=0}^{+\infty} W_n(\pmb{x}, \pmb{x}';\mu) \sigma^n(\pmb{x}, \pmb{x}').
\end{align} The so-called Hadamard coefficients \( V_n(\pmb{x}, \pmb{x}') \) can be  derived by integrating the recursion relations along the geodesic connecting \( \pmb{x} \) to \( \pmb{x}' \)~\cite{Wald:1978pj,Wald:1995yp,Moretti:2001qh, Decanini:2005eg, Moretti:2021pzz}. See Appendix \ref{AppA}. 

 The first coefficient, \( W_0(\pmb{x}, \pmb{x}';\mu) \), is not constrained by the recursion relations \eqref{delftn}, \eqref{Vrecur}, and \eqref{Wrecur} in the Appendix \ref{AppA}. This lack of uniqueness  of $W_0(\pmb{x}, \pmb{x}';\mu)$  corresponds to quantum state dependence in the biscalar \( W(\pmb{x}, \pmb{x}';\mu) \).  Once $W_0(\pmb{x}, \pmb{x}';\mu)$ is specified,  the coefficients \( W_n(\pmb{x}, \pmb{x}';\mu) \) for \( n \geq 1 \) are uniquely determined by the recursion relations \eqref{Vrecur} and \eqref{Wrecur} in Appendix \ref{AppA}. One of our main points
is to specify $W(\pmb{x}, \pmb{x}';\mu)$ including  $W_0(\pmb{x}, \pmb{x}';\mu)$ by using the mode sum expression of the Wightman function in \eqref{G+xx} and \eqref{regpart0}.

Let us recall the expanded form of the first two Hadamard coefficients~\cite{Decanini:2005eg}:
\begin{align} \label{hadamardcoeff}
V_0 (\pmb{x}, \pmb{x}') &= -1 - \frac{1}{12} \mathcal{R}_{\alpha \beta} ({\pmb x})\nabla_{\pmb x}^{\alpha} \sigma \nabla^{\beta}_{\pmb x} \sigma + \mathcal{O}(\sigma^{3/2}), \nonumber \\
V_1 (\pmb{x}, \pmb{x}') &= - \frac{m_0^2}{2} - \frac{\mathcal{R}({\pmb x})}{4} \left(2\xi - \frac{1}{3}\right) + \mathcal{O}(\sigma^{1/2}).
\end{align}

In our supersymmetric background \eqref{ourmet}, the biscalar function \( V(\pmb{x}, \pmb{x}') \) is given by:
\begin{align}\label{Vxxp}
V(\pmb{x}, \pmb{x}') &= -1 - \frac{b^2 \beta^2}{4} (\delta x^2 - \delta t^2) + \cdots \nonumber \\
&\quad + \left[ - \frac{b^2}{4} \beta (1 - 2\beta) (\delta x^2 - \delta t^2) + \cdots \right] e^{-bx} \nonumber \\
&\quad + \mathcal{O}(e^{-2bx}),
\end{align}
where the ellipses \(\cdots\) represent higher-order terms in \(\delta t \equiv t' - t\) and \(\delta x \equiv x' - x\). Here, we have used the parameters \( b \) and \( \beta \) instead of \( m_0 \) and \( \xi \) by applying the relations \( ab = \frac{m_0}{2 \xi} \) and \( \beta \equiv 2\xi \).

By employing the series expansion of the hypergeometric function \( F(A,B\,;\,C\, \mid\, -z) = 1 - \frac{AB}{C} z + \mathcal{O}(z^{2}) \), we obtain the expanded form of the product of the mode functions in \eqref{vm} and \eqref{vp} as follows:
\begin{align}\label{vvc}
&v^{(-)}_{k}({\pmb x}) (v^{(-)}_{k}({\pmb x'}))^{*}
\nonumber\\
&= e^{-i\omega_{k}(t-t') + i k (x-x')}\left[ 1 + \frac{b\beta(1-2\beta)}{b^{2}+4k^{2}} \left( \left(  b + 2ik \right) e^{-bx}
+\left(  b - 2ik \right) e^{-bx'} \right)+ \cdots \right],
\nonumber\\
&v^{(+)}_{k}({\pmb x}) (v^{(+)}_{k}({\pmb x'}))^{*}
\nonumber\\
&= e^{-i\omega_{k}(t-t') - i k (x-x')}\left[ 1 + \frac{b\beta(1-2\beta)}{b^{2}+4k^{2}} \left( \left(  b - 2ik \right) e^{-bx}
+\left(  b + 2ik \right) e^{-bx'} \right)+ \cdots \right],
\end{align}
where the ellipses \(\cdots\) denote higher-order terms in \(e^{-bx}\) and \(e^{-bx'}\).

 Using \eqref{Synexp}, \eqref{Vxxp}, and \eqref{G+xx} together with \eqref{vvc}, we can derive the expanded form of \(G^{+}_{\rm ren} ({\pmb x}, {\pmb x}';\mu)\) in \eqref{regpart0}.
In this paper, we focus on the simplest case with \(\beta = \frac{1}{2}\), which is characterized by the steepest slope of the effective potential illustrated in Figure \ref{Fig.1}.
This simplification allows us to work with analytic expressions. In this case, the renormalized Wightman function in \eqref{regpart0} is expanded as
\begin{align}\label{G+ren}
G^{+}_{\rm ren} ({\pmb x}, {\pmb x}';\mu) 
&= \frac{1}{4\pi} \left(1 + \frac{b^2}{16} \left( -(t-t')^2 + (x-x')^2 \right) \right) \ln \left( \frac{2\mu^2}{a^2b^2} \right) 
\nonumber\\
& + \frac{1}{2\pi} \left(-\gamma + \ln 2\right) 
+ \frac{b^2}{32 \pi} \left(-(t-t')^2 + (x-x')^2 \right) \left(1 - \gamma + \ln 2\right) \nonumber \\
& - \frac{1}{2\pi} \left(1 + \frac{b^2}{24} (x-x')^2 + \frac{b^2}{16} \left(-(t-t')^2 + (x-x')^2 \right) \right) 
e^{-b \frac{x + x'}{2}} + \cdots, 
\end{align}
where the ellipsis \(\cdots\) denotes higher-order terms in
\((t-t')\), \((x-x')\), and \(e^{-b \frac{x + x'}{2}}\).
Since we have considered spacelike separated two points,
the renormalized Wightman function \eqref{G+ren} is symmetric for $\pmb x$ and $\pmb x'$.

\section{Energy-momentum Tensor}\label{sec4}

In this section, we provide some computational details concerning the VEV of the energy-momentum tensor for a scalar field in our (1+1)-dimensional curved spacetime. Defining the quantum energy-momentum tensor requires particular attention, as the classical version involves products of matter fields. The distributional nature of quantum field operators complicates the naive product of fields at coincident spacetime points, resulting in significant divergences. Additionally, as discussed in Section \ref{sec3}, the issue of selecting a preferred vacuum state in curved spacetime further complicates the interpretation of the VEV of the quantum energy-momentum tensor.
To address these challenges, the algebraic approach has been developed and advocated. A key goal of this paper is to apply the algebraic approach to overcome these difficulties in calculating the VEV of the quantum energy-momentum tensor in QFTCS.

Our story begins with the energy-momentum tensor in FTCS. The bosonic part of the action \eqref{SFTCSTf} is  
\begin{align}\label{FTCSb}
S_{{\rm FTCS}} = \int d^2 x \sqrt{-g} &\Big[ -\frac{1}{2} g^{\mu\nu} \nabla_\mu\phi \nabla_\nu\phi  
- \frac{1}{2} m_0^2 \phi^2 
- \xi \mathcal{R} \phi^2\Big].
\end{align}
From this action, we read the energy-momentum tensor as 
\begin{align} \label{classicalstress}
T_{\mu\nu}
&= \nabla_{\mu}\phi\nabla_{\nu}\phi -\frac{1}{2}g_{\mu\nu}\Big[(\nabla\phi)^2 + m_{0}^2\, \phi^2\Big]  +  \xi  \Big(- \nabla_{\mu}\nabla_{\nu} + g_{\mu\nu}\nabla^{2} \Big)\phi^{2}.
\end{align}
 Based on the expression for the classical energy-momentum tensor, we now proceed to calculate the VEV of the quantum energy-momentum tensor. Using the point-splitting method and the renormalized Hadamard two-point function from~\eqref{regpart0} and~\eqref{G+ren} at the vacuum state $|0\rangle_{\rm R}^\epsilon$,  the VEV of the energy-momentum tensor is given by
\begin{equation}\label{Tmnv}
\langle T_{\mu\nu} (\pmb{x}\,;\mu) \rangle_{\rm R}^\epsilon = \lim_{\pmb{x}'\rightarrow \pmb{x}}  {\cal T}_{\mu\nu'} G^{+}_{\text{ren}} (\pmb{x}, \pmb{x}'; \mu),
\end{equation}
where the differential bi-vector in (1+1) dimensions from~\eqref{FTCSb} is expressed as
\begin{align} \label{diffBiV}
{\cal T}_{\mu\nu'} = & (1-2\xi) \partial_{\mu} \partial_{\nu'} + g_{\mu\nu'} \left[ \left(2\xi - \frac{1}{2}\right) g^{\alpha\beta'} \partial_{\alpha} \partial_{\beta'} 
- \frac{1}{2} m_0^2 + 2\xi g^{\alpha\beta} \nabla_{\alpha} \nabla_{\beta} \right] \nonumber \\
& - 2\xi \delta^{\mu'}_{\mu} \partial_{\mu'} \partial_{\nu'} + \frac{1}{4} g_{\mu\nu'} P_{x}\,.
\end{align}
Here, the last term, $P_{x} \equiv -\Box_{x} + m_0^2 + \xi \mathcal{R}$, requires further clarification~\cite{Moretti:2001qh,Moretti:2021pzz}. In traditional point-splitting renormalization of the energy-momentum tensor~\cite{Birrell:1982ix}, this term is typically omitted because it does not appear in the classical counterpart given in~\eqref{classicalstress}.

However, in the naive point-splitting approach, calculating the VEV of the energy-momentum tensor using the differential bi-vector from~\eqref{diffBiV}, while omitting the last term, can result in a failure to satisfy the covariant conservation of the VEV. 
 To restore covariant conservation, an additional regular term $g_{\mu\nu} Q$ is introduced by hand into the VEV of the energy-momentum tensor.
As emphasized in~\cite{Moretti:2001qh}, this ambiguity in the VEV conflicts with the existence of a well-defined Wick product of operators, and consequently, with the natural definition of the energy-momentum tensor operator based on this product. In~\cite{Moretti:2001qh}, a `minimal' prescription is proposed to address the shortcomings of the naive point-splitting method, where the energy-momentum tensor operator is expressed in terms of the local Wick product of field operators. This improved approach is shown to be consistent with covariant conservation.
Moreover, this prescription for including the $\frac{1}{4} g_{\mu\nu'} P_{x}$ term in~\eqref{diffBiV} is also consistently derived from the Euclidean functional approach. Thus, we adopt this prescription in our calculation of the VEV of the energy-momentum tensor in the (1+1)-dimensional background.

Despite this prescription, an intrinsic renormalization ambiguity in the mass scale $\mu$ persists in the resulting VEV expression of the energy-momentum tensor operator, manifesting as a logarithmic divergence in two dimensions in the form $\ln (\mu^{2} \sigma)$. Specifically, the subtraction term in the Hadamard renormalization takes the form of \eqref{regpart0}:
\begin{align} \label{regpart}
- \frac{1}{4 \pi} V({\pmb x}, {\pmb x}') \ln \Big[ \mu^{2} \sigma ({\pmb x}, {\pmb x}') \Big]\,.
\end{align}
Typically, $\mu$ is chosen such that the VEV of the energy-momentum tensor vanishes in the Minkowski vacuum, {\it i.e.}, ${}_{\rm M}\langle 0| T_{\mu\nu}|0 \rangle_{\rm M} = 0$.

According to our above discussion for the VEV of the quantum energy-momentum tensor, now we provide explicit calculational result of $\langle T_{\mu\nu} (\pmb{x}\,;\mu) \rangle_{\rm R}^\epsilon$ in \eqref{Tmnv}. Inserting \eqref{G+ren} into \eqref{Tmnv}, we obtain  
\begin{align}    \label{Ttt}
\langle T_{tt}\rangle_{\rm R}^\epsilon  &=   \frac{b^{2}}{16\pi} \Big[\frac{1}{2}-\gamma + 
   \ln 2 +\frac{1}{2} \ln \Big(\frac{2 \mu^2}{a^2b^2}  \Big) \Big] \Big[1- 2 e^{- bx}\Big]  
\nonumber\\
&~~~  -\frac{b^2}{12\pi} e^{- bx} +  {\cal O} \Big(e^{-2 b x}\Big) \,,  \\ \label{Txx}
\langle T_{xx}\rangle_{\rm R}^\epsilon  &=   \frac{b^{2}}{16\pi} \Big[- \frac{1}{2} + \gamma - 
   \ln 2 -\frac{1}{2} \ln \Big(\frac{2 \mu^2}{a^2b^2}  \Big) \Big]   \Big[1- 2 e^{- bx}\Big] 
+  {\cal O} \Big(e^{-2 b x}\Big) \,,   \\
\langle T_{tx} \rangle_{\rm R}^\epsilon  &= \langle T_{xt}  \rangle_{\rm R}^\epsilon ={\cal O} \Big(e^{-2 b x}\Big) \,. \label{Ttx}   
\end{align}
Since $\mathcal{R} = 2a^2b^2 e^{-bx}$(see footnote \ref{abouta}), the expansion presented above can be understood as an expansion in the weak curvature limit as $x \to \infty$.

Before advancing further, it is desirable to verify the covariant conservation of the VEV of the energy-momentum tensor. As demonstrated by Moretti in~\cite{Moretti:2001qh}, the conventional point-splitting method is improved by incorporating the \( P_x \) term in~\eqref{diffBiV}, preserving covariant conservation under the assumption that a global Hadamard state exists. However, since the existence of a global Hadamard state within our background geometry is not expected  because of the naked null singularity,
it is worthwhile to examine the covariant conservation of the VEV of the energy-momentum tensor explicitly. This can be confirmed up to the order \( \mathcal{O}(e^{-2b x}) \). Specifically, we find that
\begin{equation} \label{delTmn}
\nabla_{\mu} \langle T^{\mu\nu} \rangle_{\mathrm{R}}^\epsilon = 0,
\end{equation}
holds up to terms of order $ e^{-2b x} $. Note that the choice of the renormalization mass scale $\mu$ is irrelevant to the covariant conservation \eqref{delTmn}.

As \( x \rightarrow \infty \), the state \( |0\rangle_{\mathrm{R}}^\epsilon \) asymptotically approaches   Minkowski vacuum \( |0\rangle_{\mathrm{M}} \). This is evident since, by taking the limit \( x \to \infty \) in the metric~\eqref{ourmet}, we recover  Minkowski metric:
\begin{align}\label{ourMin}
ds^2 = \frac{1}{a^2} \left(-dt^2 + dx^2 \right).
\end{align}
It is apparent that our expressions for the VEV of the energy-momentum tensor, as given in equations \eqref{Ttt}, \eqref{Txx}, and \eqref{Ttx}, do not vanish in this limit or in  Minkowski spacetime. However, this  is associated with the freedom to choose the mass scale \( \mu \). By selecting an appropriate mass scale \( \mu \), the scheme-dependent part  in the renormalization procedure can be removed. As discussed previously, the parameter \( \mu \) is determined such that the VEV of the energy-momentum tensor vanishes in Minkowski spacetime.

In this Minkowski metric, we have\begin{align}
\langle T_{tt}\rangle_{\rm M}  &=  
\frac{m_0^2}{4\pi a^2 } \left[\frac{1}{2}-\gamma 
+ \ln 2 
+ \frac{1}{2}\ln\!\left(\frac{\mu^2}{2 m_0^2}\right) \right],
\nonumber\\[4pt]
\langle T_{xx}\rangle_{\rm M}  &=  
\frac{m_0^2}{4\pi a^2} \left[-\frac{1}{2}+ \gamma 
- \ln 2 
- \frac{1}{2}\ln\!\left(\frac{\mu^2}{2 m_0^2}\right) \right],
\nonumber\\[4pt]
\langle T_{tx}\rangle_{\rm M} &= \langle T_{xt}\rangle_{\rm M} = 0.
\end{align}
To ensure that \( \langle T_{\mu\nu}\rangle_{\rm M} = 0 \), we choose the renormalization scale \( \mu \) as
\begin{align}\label{rensc}
\mu = \frac{m_0}{\sqrt{2}}\, e^{\gamma - \frac{1}{2}}.
\end{align}
Finally, the VEV of the renormalized energy-momentum tensor is given by 
\begin{align}
\langle T_{tt}\rangle_{\rm R}^\epsilon  &=-  
  \frac{b^{2}}{12\pi}    e^{-bx} +   {\cal O} \big(e^{-2 b x}\big) \,,  \nonumber \\ \label{Txx2}
\langle T_{xx}\rangle_{\rm R}^\epsilon  &=    {\cal O} \big(e^{-2 b x}\big) \,,   \\
\langle T_{tx} \rangle_{\rm R}^\epsilon  &= \langle T_{xt}  \rangle_{\rm R}^\epsilon ={\cal O} \big(e^{-2 b x}\big) \nonumber\,\,.
\end{align}
The non-vanishing quantities in the above result represent the quantum effects of the state 
$|0\rangle_{{\rm R}}^\epsilon$ , which reflects the fact that the classical lowest energy must be zero.
Note that using the result in \eqref{Txx2} and recalling the expression of the Ricci scalar in footnote \ref{abouta}, we obtain the trace of the VEV,  
\begin{align}
g^{\mu\nu}\langle T_{\mu\nu} \rangle_{\rm R}^\epsilon = \frac{1}{12\pi} a^2 b^2 e^{-bx} +{\cal O} \big(e^{-2 b x}\big) = \frac{1}{24\pi} \mathcal{R} +{\cal O} \big(e^{-2 b x} \big).
\end{align}

At first glance, it appears to coincide with the trace anomaly; however, it is evident that our model is not conformally symmetric, so further clarification is required. In general, the trace of the VEV of the energy-momentum tensor in two dimensions (see, for example, \cite{Decanini:2005eg}) is given by
\begin{align}\label{traceT}
g^{\mu\nu}\langle T_{\mu\nu} (\pmb{x}; \mu) \rangle =  (\xi \Box - m_0^2) G^{+}_{\text{ren}} (\pmb{x}, \pmb{x}; \mu) + \frac{1}{24 \pi} \mathcal{R}.
\end{align}
In the special case of \(\xi = 0\) and \(m_0 = 0\), we have a two-dimensional conformal theory, which reproduces the conformal anomaly as
\[
g^{\mu\nu}\langle T_{\mu\nu} (\pmb{x}; \mu) \rangle_{\rm R}^\epsilon = \frac{1}{24 \pi} \mathcal{R}.
\]
In our case, since \(\xi = \frac{1}{4}\) and \(m_0 = 2ab \xi\), the first term on the right-hand side of \eqref{traceT} appears to be nonvanishing. However, it actually vanishes up to the relevant order because
\begin{align}
\Box G^{+}_{\text{ren}} (\pmb{x}, \pmb{x}; \mu) = a^2 b^2 G^{+}_{\text{ren}} (\pmb{x}, \pmb{x}; \mu) + {\cal O} \big(e^{-2 b x} \big),
\end{align}
where \(G^{+}_{\text{ren}} (\pmb{x}, \pmb{x}; \mu) = - \frac{1}{2\pi} e^{-bx} + {\cal O} \big(e^{-2 b x} \big)\) as given in \eqref{G+ren} with \( \mu = \frac{m_0}{\sqrt{2}}\, e^{\gamma - \frac{1}{2}}\). At this stage, the nonvanishing trace of the VEV cannot be interpreted as a conformal anomaly but coincides with the same expression accidentally.

Comments on the $\langle T_{\mu\nu}\rangle_{{\rm R}}^\epsilon$-component in \eqref{Txx2} are  in order. Within the context of Hadamard renormalization, after eliminating the divergence from the Wightman function, the remaining term can be interpreted as the effect of a quantum state. Accordingly, using this prescription, we interpret the result $\langle T_{\mu\nu}\rangle_{\rm R}^\epsilon$ in \eqref{Txx2} as representing the effect of the quantum state $|0\rangle_{{\rm R}}^\epsilon$. As discussed in section \ref{sec3}, the Hilbert space containing the state $|0\rangle_{{\rm R}}^\epsilon$ is relevant for an observer located slightly out of the right asymptotic region. It should be noted that this Hilbert space is not to be interpreted as a local one\footnote{ It is well known that there does not exist micro states  for a local region. See \cite{Haag:1992hx,Witten:2021jzq,Witten:2023qsv} for example.}; however, for the local observer, this Hilbert space is perceived as a ``global'' one. In other words, the local observer mimics the global plane wave quantization through R-quantization, as discussed in section \ref{sec3}, by restricting a valid range of $\omega \ge b \beta $ without imposing any constraints on position.

Now, let us consider the physical implications of $\langle T_{tt}\rangle_{\rm R}^\epsilon < 0$ in the aforementioned setup. To facilitate the discussion, it is useful to recall a similar result in the context of Rindler spacetime.  
If $\langle T_{tt} \rangle_{\rm M}$, the VEV of $T_{tt}$ for the Minkowski vacuum $|0\rangle_{\rm M}$ obtained via canonical quantization, is set to zero by the renormalization condition, then $\langle T_{tt} \rangle_{\rm Rindler}$, the VEV of $T_{tt}$ for the Rindler vacuum $|0\rangle_{\rm Rindler}$, becomes negative. This implies that the energy density of the Rindler vacuum state is lower than that of the Minkowski vacuum, which is related to the Unruh effect.
In the context of canonical quantization, since the Fock space $\mathcal{F}_{\rm M}$ constructed from the Minkowski vacuum $|0\rangle_{\rm M}$ is not unitarily related to the Fock space $\mathcal{F}_{\rm Rindler}$, such a comparison of the two energy densities is not conceptually appropriate. Therefore, we need to adopt the algebraic description, in which these two unitarily inequivalent Fock spaces are regarded as different representations of the same field algebra~\cite{Ho:2022omx}. 
Turning to our model, we would like to interpret the fact that $\langle T_{tt}\rangle_{\rm R}^\epsilon < 0$ as follows: In analogy with the Rindler case discussed above, we interpret $\langle T_{tt}\rangle_{\rm R}^\epsilon < 0$ as indicating that an observer located slightly out of the right asymptotic region would detect a thermal-like particle distribution for the field in the right Minkowski vacuum. We leave the clarification of this analogous interpretation for future work.

\section{Quantum Effects in IFT}\label{sec5}

In this section, we elaborate the description given in \cite{Kwon:2022fhv,Kwon:2021flc} to elucidate quantum effects within QIFT. 

\subsection{Quantum states in QIFT}\label{subsec5.1}

As shown in subsection \ref{subsec2.3}, while the L-vacuum $|0\rangle_{\rm L}$ defined in \(\eqref{leftvacuum}\) is suitable for the left region \(x \to -\infty\), it is not an appropriate candidate for the vacuum of the scalar field $\phi$ in the right region \(x \to \infty\).
Since (massless) particles of the energy $0\le \omega \le b \beta $ in the left region do not exist in the right asymptotic region,  the L-vacuum  cannot be an appropriate vacuum for the quantum field in the right region, $x\to + \infty $.
For similar reasons, the R-vacuum $|0\rangle_{\rm R}$ is not a good candidate for a vacuum state in the left asymptotic region. As discussed in~\cite{Ho:2022omx}, within the L-quantization scheme, the Fock space \( \mathcal{F}_{\mathrm{L}} \) is constructed by massless particles, whereas the R-quantization scheme yields a Fock space \( \mathcal{F}_{\mathrm{R}} \) consisting of massive particles. The Fock space \( \mathcal{F}_{\mathrm{L}} \) provides a good representation in the region where \( x \to -\infty \), while \( \mathcal{F}_{\mathrm{R}} \) offers a valid description for \( x \to \infty \). These two Fock spaces cannot be connected by a unitary transformation, making the existence of a global Hadamard state implausible.

As alluded before, in the context of FTCS, L-quantization is conceptually ill-defined due to the presence of a naked singularity as $x \to -\infty$. However, from the perspective of IFT, constructing $\mathcal{F}_{{\rm L}}$ through L-quantization presents no such difficulty. In the following, we will discuss the bubble expansion of the Higgs condensation, considering both the L-quantization and the R-quantization.

\subsection{Two-point function and energy-momentum tensor in QIFT}\label{subsec5.2}
Specifically, we reinterpret the descriptions provided in sections \ref{sec3} and \ref{sec4} within the framework of QIFT, applying the concepts of R-quantization discussed in subsection \ref{subsec2.3}. While we were able to construct the vacuum state $|0\rangle_{{\rm R}}$ in the right asymptotic region within the context of QIFT, as discussed in subsection \ref{subsec2.3}, a method for renormalizing  divergences appearing in the two-point function with respect to $|0\rangle_{{\rm R}}$ has not yet been established. In this regard, we propose a renormalization procedure in QIFT by extending the classical relationship between FTCS and IFT to the quantum domain.

As is well-known in QFTCS, the divergence in the coincident limit $\pmb{x}' \to \pmb{x}$ of the two-point function \(\omega_{\rm C}(\phi(\pmb{x})\phi(\pmb{x}'))\) appears for an algebraic state \(\omega_{\rm C}\) in curved spacetime. This divergence structure is universal. One way to address this divergence is through the Hadamard regularization method. In curved spacetimes, Hadamard regularization is characterized by a specific short-distance behavior of the two-point function, where the short-distance divergence structure is expressed by particular combinations of the Synge function. Based on our previous proposal for QIFT~\cite{Kwon:2022fhv, Ho:2022omx}, we transcribe  these conditions into the QIFT framework. In this context, \( \omega_{\rm R}^{\epsilon} \) discussed in subsection \ref{subsec3.1} is referred to as the Hadamard state in QIFT.
Specifically, we propose that the renormalized two-point function in QIFT is identified with $G^{+}_{\text{ren}} (\pmb{x}, \pmb{x}'; \mu) $ as defined in \eqref{regpart0}, by choosing the same subtracting term of  \(\frac{1}{4 \pi} V(\pmb{x}, \pmb{x}') \ln \left[ \mu^2\sigma (\pmb{x}, \pmb{x}') \right]\).

Now, let us consider the energy-momentum tensor in IFT. Defining the classical energy-momentum tensor in IFT as a conserved current is problematic due to the lack of Poincar\'e symmetry in the IFT action. However, although the theory of gravity also lacks  Poincar\'e symmetry, we can still define the energy-momentum tensor as a source of the gravitational field, which satisfies the general covariance, by varying the gravity Lagrangian with respect to the metric tensor. Since our IFT model possesses time translation symmetry, the energy of the system is conserved. Therefore, the $(tt)$-component of the energy-momentum tensor can be constructed canonically. However, there is no explicit criterion for determining the other components of the energy-momentum tensor. 
Under these circumstances,
we have proposed a way to define the energy-momentum tensor in IFT, using the conversion relation between  IFT and  FTCS in (1+1)-dimensions~\cite{Kwon:2022fhv}.
  Although this proposal is somewhat restrictive, it enables us to perform explicit calculations and make physical interpretations within the context of IFT.
Then, the result for the VEV of the energy-momentum tensor in FTCS from Section \ref{sec4} can be directly transcribed to the VEV of the energy-momentum tensor in QIFT.
Based on this, we aim to interpret our results in the context of Higgs condensate bubble expansion.

\subsection{Higgs condensate bubble expansion}\label{subsec5.3}

The dynamics of bubble expansion during the first-order electroweak phase transition in the early universe have been extensively studied~\cite{Coleman:1977py,Callan:1977pt,Linde:1981zj}. In particular, in the context of the thick bubble wall model, a Lagrangian with a position-dependent mass was introduced and analyzed~\cite{Polyakov:1974ek,Dine:1992vs, Moore:1995si,Moore:1995si,Moore:1995ua,Bodeker:2009qy,Bodeker:2017cim,Azatov:2020ufh, Kubota:2024wgx}. Building upon the discussions in subsections~\ref{subsec5.1} and \ref{subsec5.2}, we now focus on the quantum effects influencing the expansion of the Higgs condensate bubble.
 We utilize the model specified in \eqref{IFTact}, which was also employed in~\cite{Arnold:1993wc, Moore:1995si, Moore:1995si, Moore:1995ua, Bodeker:2009qy, Bodeker:2017cim, Azatov:2023xem, Kubota:2024wgx}.

It has been argued that various frictional effects arise from the interaction with the plasma in the symmetric phase, along with a driving force due to the potential difference between the false and true vacua at finite temperature~\cite{Linde:1981zj}. Complementing these findings, Kubota employed the spectral functions   at zero temperature  to  investigate Green functions for the bubble wall model~\cite{Kubota:2024wgx}. 
In this subsection, we propose the frictional effects acting on the bubble wall expansion, those arising from the quantum vacuum state at zero temperature, in the context of the {\it quantum} energy-momentum tensor in the IFT framework.

The total pressure exerted on the bubble wall can be decomposed as
\begin{align}
    P_{\rm tot} 
    = -\Delta V + \Delta  P,
\end{align}
where $\Delta V$ and $\Delta P$ represent the differences in potential and pressure, respectively, between the symmetric phase and the Higgs phase.
Here, $\Delta V$ generates the driving force for the expansion of the bubble wall, while $\Delta P$, with a positive sign, represents the frictional force opposing the expanding bubble wall.
In the context of our effective IFT model for bubble wall expansion, since we have adopted the bubble wall rest frame, the driving force $\Delta V$ becomes irrelevant, while $\Delta P$ needs to be taken into account.
Note that $\Delta P$ includes both classical and quantum effects, {\it i.e.,}
\begin{align}
    \Delta P = \Delta P_{\rm classical} + \Delta P_{\rm quantum}.
\end{align}
In this expression, the thermal effect would appear in the pressure difference. However, in our zero-temperature model, it turns out that $\Delta P_{\rm classical} = 0$ because our classical vacuum satisfies $\phi_{\rm classical} = 0$, which implies $T_{\mu \nu}^{\rm classical} = 0$. (See subsection 2.5 of \cite{Ho:2022omx}.)

As discussed in subsection~\ref{subsec5.2}, the {\it quantum} energy-momentum tensor in IFT, $\langle T_{\mu \nu} \rangle^\epsilon_{\rm IFT}$, is identified with $\langle T_{\mu \nu} \rangle^\epsilon_{\rm R}$ as given in \eqref{Txx2}, where $\epsilon$ represents the bubble wall region {\it close} to the Higgs phase. It is important to emphasize that the term {\it close} to the Higgs phase refers to a location slightly out of the right asymptotic region (see Figure \ref{Fig.1}).
Following the prescription provided in \cite{Linde:1981zj}, the frictional force can be determined by the pressure difference:  
\begin{align}\label{DPe}
\Delta P_\epsilon = P_{{\rm \epsilon}} - P_{{\rm H}} = \langle T_{xx}\rangle^\epsilon_{{\rm IFT}} -  \langle T_{xx}\rangle^{{\rm H}}_{{\rm IFT}},
\end{align}
where $\langle T_{xx}\rangle^{{\rm H}}_{{\rm IFT}}$ represents the VEV of the $xx$-component of the IFT energy-momentum tensor calculated in the Higgs vacuum. Here, according to our conversion proposal in subsection \ref{subsec5.2}, $\langle T_{xx}\rangle^{{\rm H}}_{{\rm IFT}} = \langle T_{xx}\rangle_{{\rm R(M)}} = 0$.
As a result,  $\Delta P_\epsilon$ in \eqref{DPe} is read from \eqref{Txx2} as
\begin{align}
\Delta P_\epsilon=       {\cal O} \big(e^{-2 b x}\big)  \,.
\end{align}
From the point of view the inhomogeneous field theory, the quantum vacuum 
yields an Unruh--like contribution as the dominant term, scaling as 
$\mathrm{e}^{-b x}$. 
However, any pressure---whether tending to hinder or to assist the bubble--wall 
expansion---does not arise at this leading order. 
Accordingly, the influence of the pressure difference $\Delta P$ enters only at 
the next--to--leading order, suppressed by at least $\mathrm{e}^{-2 b x}$.

\section{Conclusions}

In this paper, we examined a free scalar field \( \phi \) in a supersymmetric background metric given by \eqref{ourmet} in (1+1) dimensions. After deriving the renormalized Wightman function, we explicitly obtained the covariantly conserved \textit{quantum} energy-momentum tensor up to \( {\cal O} \big(e^{-2 b x}\big) \). Based on these results, we have shown the existence of the Unrhu-like quantum effects in the leading order $(e^{- b x})$on the expansion of the bubble wall during the electroweak phase transition in the early universe.

Our results show that the Hadamard renormalization scheme, which has been successfully applied in traditional QFTCS, also provides an effective method for addressing the renormalization of the Wightman function in QIFT. This accomplishment shows  the consistency of our proposal in constructing meaningful physical quantities in QIFT.

The entire calculation in our paper follows a conceptually consistent procedure by introducing an algebraic state \( \omega_{{\rm R}}^\epsilon \) (or \( |0\rangle_{{\rm R}}^\epsilon \)), which is locally Hadamard. Using this algebraic state, we have shown that \( \langle T_{tt}\rangle_{\rm R}^\epsilon < 0 \) and \( \langle T_{xx}\rangle_{\rm R}^\epsilon = 0 \) in the leading order $e^{- b x}$. From the result \( \langle T_{tt}\rangle_{\rm R}^\epsilon < 0 \), we inferred the existence of an Unruh-like effect for an observer located at \( x_\epsilon \).

There are various future directions to pursue. One possibility is to extend our computations to the left asymptotic region, where a naked singularity exists. By introducing a suitable regularization parameter to handle the singularity, we can pursue similar calculations in this context. Another direction involves extending our method to a free fermion field in the framework of IFT~\cite{HKPYfermion}. Additionally, exploring finite temperature effects in QIFT could provide insights relevant to the expansion of bubble walls in the early universe. Finally, investigating higher-dimensional QIFT would be of particular interest. In this case, however, the use of the Synge function in Hadamard regularization in the IFT context becomes nontrivial, as the (1+1)-dimensional conversion rule between FTCS and IFT does not apply.

\section*{Acknowledgments}
We appreciate conversations and discussions with Sang-A Park, as well as the valuable email correspondence with Takahiro Kubota. This work was supported by the National Research Foundation of Korea (NRF) grant with grant number RS-2023-00208047 (J.H.), RS-2019-NR040081, RS-2023-00249608 (O.K.), NRF-2022R1F1A1076172, NRF-2021R1A2C1003644 (S.-H.Y.), and supported by Basic Science Research Program through the NRF funded by the Ministry of Education NRF-2020R1A6A1A03047877 (J.H. and S.-H.Y.).

\newpage

\begin{center} {\Large \bf Appendix}
\end{center}

\begin{appendix}

\section{Hadamard Expansion of Two-point Function} \label{AppA}

In this Appendix, we summarize the Hadamard expansion of the two-point function in (1+1) dimensions. We start with the Wightman function with positive frequencies, which can be represented by the biscalar functions $V(\pmb{x}, \pmb{x}')$ and $W(\pmb{x}, \pmb{x}'; \mu^2)$ in \eqref{VWnxxp} as
\begin{align}\label{GFxxp}
G^{+}(\pmb{x}, \pmb{x}') = \frac{1}{4\pi} \left[ V(\pmb{x}, \pmb{x}') \ln \big(\mu^2 \sigma (\pmb{x}, \pmb{x}') \big) + W(\pmb{x}, \pmb{x}'; \mu^2) \right],
\end{align}
where we consider $\sigma(\pmb{x}, \pmb{x}')$ for space-like separated points for simplicity.
Inserting the relation in \eqref{GFxxp} into the Green's equation for the Wightman function, we obtain
\begin{align}
(-\Box_x + m_0^2 + \xi \mathcal{R}) G^+(\pmb{x}, \pmb{x}') = \frac{1}{\sqrt{-g}} \delta(t - t') \delta(x - x'),
\end{align}
and expanding in terms of the Synge function $\sigma(\pmb{x}, \pmb{x}')$, we obtain three relations~\cite{Decanini:2005eg}:
\begin{align}
&  \frac{V_0}{4 \pi} \Box \ln \sigma + \frac{1}{2 \pi} \left(\nabla_\mu V_0 \nabla^\mu \sigma \right) \sigma^{-1} =- \frac{1}{\sqrt{-g}} \delta(t - t') \delta(x - x'),
\label{delftn}\\
&  (- \Box + m_0^2 + \xi \mathcal{R}) V_n - 2(n+1)^2 V_{n+1} - 2 (n+1) \nabla_\mu V_{n+1} \nabla^\mu \sigma 
\nonumber\\
&~~ + 2 (n+1) V_{n+1} \frac{1}{\sqrt{\Delta}} \nabla_\mu \sqrt{\Delta} \nabla^\mu \sigma = 0,
\label{Vrecur}\\
&  (- \Box + m_0^2 + \xi \mathcal{R}) W_n - 2(n+1)^2 W_{n+1} - 2 (n+1) \nabla_\mu W_{n+1} \nabla^\mu \sigma
\nonumber\\
& ~~+ 2 (n+1) W_{n+1} \frac{1}{\sqrt{\Delta}} \nabla_\mu \sqrt{\Delta} \nabla^\mu \sigma - 4(n+1) V_{n+1} - 2 \nabla_\mu V_{n+1} \nabla^\mu \sigma
\nonumber\\
& ~~+ 2 V_{n+1} \frac{1}{\sqrt{\Delta}} \nabla_\mu \sqrt{\Delta} \nabla^\mu \sigma = 0
\label{Wrecur}
\end{align}
with the boundary condition
\begin{align}
V_0(\pmb{x}, \pmb{x}') = - \Delta^{\frac{1}{2}}(\pmb{x}, \pmb{x}').
\end{align}
In deriving the relations \eqref{delftn} - \eqref{Vrecur}, one can use several properties of the Synge function and the van Vleck-Morette determinant given in \eqref{syngediffeq1} - \eqref{Dxxp}. In the coincident limit $\pmb{x}' \to \pmb{x}$, the geometry reduces to Minkowski space, i.e., $g_{\mu\nu} \to \eta_{\mu\nu}$ and $V_0(\pmb{x}, \pmb{x}') \to -1$. Then, the relation reduces to
\begin{align}
\frac{1}{4 \pi} \partial_\mu \partial^\mu \ln \sigma = \delta(t - t') \delta(x - x'),
\end{align}
where $\sigma = \frac{1}{2} \left( - (t - t')^2 + (x - x')^2 \right)$. This corresponds to the fact that the singular part of the Wightman function in Minkowski space is given by $G_{\text{sing}}^+ = -\frac{1}{4 \pi} \ln \sigma$.

\section{Order Estimate of $\Phi_{\epsilon}$} \label{AppB}

In this Appendix, we present an argument to estimate the order of $\Phi_{\epsilon}$ in~\eqref{Phieps}.  
To proceed, let us consider a ``position-dependent'' dispersion relation given by
\begin{equation} \label{dispersion}
\omega^{2}(x) = k^{2} + m^{2}(x)\,.
\end{equation}
Under the condition $\beta = \frac{1}{2} = 2\xi$, we expand the mass function around the point $x = x_{\epsilon}$, which is located near $x = \infty$. This expansion yields
\begin{equation} \label{mass-expansion}
m^{2}(x) = \left( \frac{m_{0}}{a} \right)^{2} + e^{-bx} \left[ \frac{b^{2}}{2a} - 2\frac{m_{0}^{2}}{a^{3}} \right] + e^{-2bx} \left[ -\left( \frac{b}{a} \right)^{2} + 3\frac{m_{0}^{2}}{a^{4}} \right] + {\cal O}\left( e^{-3bx} \right)\,.
\end{equation}
In our case, the relation $ab = 2m_{0}$ holds, which leads to
\begin{equation*}
\frac{b^{2}}{2a} - 2\frac{m_{0}^{2}}{a^{3}} = 0\,.
\end{equation*}
Consequently, the expansion of the mass function simplifies to
\begin{equation} \label{mass-simplified}
m^{2}(x) = (b\beta)^{2} - \frac{b^{2}}{4a^{2}}\, e^{-2bx} + {\cal O}\left( e^{-3bx} \right)\,.
\end{equation}
Now, using the dispersion relation, we deduce the relation between $x_{\epsilon}$ and $\epsilon$. As illustrated in Fig.~\ref{Fig.1}, the energy at the position $x = x_{\epsilon}$ for the lowest momentum ({\it i.e.}, $k = 0$) is given by
\begin{equation*}
\omega = b\beta - \epsilon \qquad (\epsilon \ll b\beta)\,.
\end{equation*}
Substituting this into the dispersion relation~\eqref{dispersion}, we obtain
\begin{equation} \label{omega-epsilon}
\omega^{2} = (b\beta - \epsilon)^{2} = m^{2}(x_{\epsilon}) = \left( \frac{b}{2} \right)^{2} - \frac{b^{2}}{4a^{2}}\, e^{-2bx_{\epsilon}} + {\cal O}\left( e^{-3bx_{\epsilon}} \right)\,.
\end{equation}
Expanding the left-hand side for small $\epsilon$, and recalling $\beta = \frac{1}{2}$, we find
\begin{equation} \label{xeps-relation}
\frac{1}{2b\beta}\epsilon = \frac{1}{b} \epsilon = \frac{b}{4a^{2}} e^{-2bx_{\epsilon}} + {\cal O}\left( e^{-3bx_{\epsilon}} \right)\,.
\end{equation}
Recalling $W \sim e^{-i(\omega t \pm kx)} \sim {\cal O}(1)$ in~\eqref{Phieps}, we can estimate the order of the integral term in $\Phi_{\epsilon}$ as
\begin{equation} \label{integral-estimate}
\int^{b\beta}_{b\beta - \epsilon} d\omega \left( \cdots \right) \simeq \epsilon \times {\cal O}(1) \simeq {\cal O}\left( e^{-2bx_{\epsilon}} \right)\,.
\end{equation}
Since this term is of higher order compared to the terms of interest, which are up to ${\cal O}\left( e^{-bx_{\epsilon}} \right)$ in $\Phi_{\epsilon}$, we conclude that the integral term over the range $b\beta - \epsilon < \omega < b\beta$ in $\Phi_{\epsilon}$ can be neglected in our analysis. Consequently, the contribution of $\Phi_{\epsilon}$ to the 2-point function is ${\cal O}(e^{-2bx_{\epsilon}})$, which is negligible to the leading  order of interest in the 2-point function, ${\cal O}(e^{-bx_{\epsilon}})$.

\end{appendix}

\end{document}